\documentclass[traditabstract]{aa} 
	
\usepackage{graphicx}
\usepackage{color}
\usepackage{txfonts}

\def\MgII{Mg\,{\sc ii}}
\def\RAA{\rm \AA}

\begin{document}	

   \title{Incidence of \MgII\ absorbers towards Blazars\\ and the GRB/QSO puzzle\thanks{Based on observations collected at the European Southern Observatory, ESO, the VLT/Kueyen telescope, Paranal, Chile, in the framework of programs 080.A-0276 and 081.A-0193.} }
    \titlerunning{\MgII\ absorbers towards Blazars}	
	
   \author{Jacqueline Bergeron	
          \inst{1}	
          \and	
          Patrick Boiss\'e\inst{1}	
          \and	
          Brice M\'enard\inst{2}
         }	
	
   \institute{Institut d'Astrophysique de Paris (IAP), CNRS-UPMC,	
              98bis Boulevard Arago, F-75014 Paris, France\\	
              \email{bergeron@iap.fr}	
         \and	Canadian Institute for Theoretical Astrophysics, University of 
              Toronto, 60 St. George Street, Toronto, Ontario, M55 3H8, Canada 	
             }	
	
   \date{Received ; accepted }	
	
 	
   \abstract {
In order to investigate the origin of the excess of strong Mg\,{\sc ii} systems towards 
GRB afterglows as compared to QSO sightlines, we have measured the incidence of 
\MgII\ absorbers
towards a third class of objects: the Blazars. This class includes the BL Lac object 
population for which a tentative excess of Mg\,{\sc ii} systems had already been reported.
We observed with FORS1 at the ESO-VLT 42 Blazars with an emission redshift 
$0.8<z_{\rm em}<1.9$, to which we added the three high $z$ northern objects belonging to the 
1Jy BL Lac sample. We detect 32 Mg\,{\sc ii} absorbers in the redshift range 0.35-1.45, 
leading to an excess in the incidence of Mg\,{\sc ii} absorbers compared to that measured 
towards QSOs by a factor $\sim 2$, detected at 3$\sigma$. The amplitude of the effect is 
similar to that found along GRB sightlines. Our analysis provides a new piece of evidence 
that the observed incidence of \MgII\ absorbers might depend on the type of background source. 
In front of Blazars, the excess is apparent for both 'strong' ($w_{\rm r}$(2796)$ 
> 1.0$~\AA) and weaker ($0.3 < w_{\rm r}$(2796)$ < 1.0$~\AA) \MgII\ systems. The 
dependence on velocity separation with respect to the background Blazars indicates, 
at the $\sim1.5\sigma$ level, a potential excess for $\beta\equiv v/c\sim0.1$. 
We show that biases involving dust extinction or gravitational amplification are not likely 
to notably affect the incidence of Mg\,{\sc ii} systems towards Blazars. Finally we discuss 
the physical conditions required for these absorbers to be gas entrained by the powerful 
Blazar jets. More realistic numerical modelling of jet-ambient gas interaction 
is required to reach any firm conclusions as well as repeat observations at high 
spectral resolution of strong Mg\,{\sc ii} absorbers towards Blazars in both high and 
low states. }


   \keywords{BL Lacertae objects: general -- quasars: absorption lines}	
	
   \maketitle	
	
\section{Introduction}	
	
In the 1980's, the observation of large QSO samples allowed 
Young et al. (\cite{young}) and a decade later Steidel \& Sargent 
(\cite{steidel}) to establish that C\,{\sc iv} and  
Mg\,{\sc ii} absorption systems arise in intervening material, randomly	
distributed along the QSO sightlines.  This conclusion was based on two main	
arguments: i) the number of systems detected in individual sightlines is	
Poissonian; ii) the distribution of systems with respect to comoving	
coordinates is uniform. Their Mg\,{\sc ii}  sample did not reveal any 
cosmological evolution. 
	
To firmly establish the nature of the lower redshift  Mg\,{\sc ii} absorption 		
systems, searches for galaxies associated with these absorptions have been 		
conducted since the late 1980's. 	
 For Mg\,{\sc ii} systems at $z\lesssim 1$, results of these identification 	
surveys revealed that normal field galaxies with 	
luminosities $L>0.1L^{\star}$ are all potentially Mg\,{\sc ii} absorbers,  		
and the mean extent of their gaseous halos is $\sim$65 kpc, assuming 
$H_0=70$ km s$^{-1}$ Mpc$^{-1}$ (Bergeron \& Boiss\'e \cite{bergeron91};		
Steidel et al. \cite{steidel94}; Steidel \cite{steidel95}). The 		
conclusion of these searches was that in studying Mg\,{\sc ii}  		
systems, we get information on the cosmic history of normal field galaxies 		
(metallicity, dust content, dynamics, star formation). 			
	
Our current understanding of absorption line systems thus implicitly assumes  	
that their properties do not depend on the nature of the background sources 	
used as probes (except at $z_{\rm abs}\simeq z_{\rm em}$). 
However, Prochter et al. (\cite{prochtera}) recently reported an excess of 
strong (rest equivalent width  $w_{\rm r}$(Mg\,{\sc ii}2796) $> 1.0$~\AA)  
Mg\,{\sc ii} absorbers towards gamma-ray bursts (GRBs) with respect to the 
incidence measured in QSO spectra. These authors found that nearly every GRB 
sightline exhibits at least one absorber ($\langle z \rangle = 1.1$) whereas the 
incidence of strong Mg\,{\sc ii} QSO absorbers is about four times smaller.
%
%
This excess is confirmed by Sudilovsky et al. (\cite{sudilovsky}) 	
for a small, homogeneous sample of GRB VLT-UVES high-resolution spectra, 	
and more recently by Tejos et al. (\cite{tejos09}) and 
Vergani et al. (\cite{vergani}). Based on their improved 
statistics, these latter authors find the excess to be smaller than 
initially reported, by a factor of $2.1 \pm 0.6$ instead of $\sim$4. 
There is no such excess for the higher redshift C\,{\sc iv} absorbers 
(small overlapping samples of Sudilovsky et al. \cite{sudilovsky} and  
Tejos et al. \cite{tejos07}), which trace 	
more diffuse and hotter gas than Mg\,{\sc ii}, have larger cross-sections 	
and, therefore, largely probe a different population than strong 
Mg\,{\sc ii} absorbers.	
	
Studies of systems towards GRBs naturally focused	
on the strongest systems, the only ones that could be easily detected and 
thus for which homogeneous results could be obtained (since they fade away 
rapidly, GRB afterglows are generally faint when observed at good spectral 
resolution). Initially, the incidence of strong Mg\,{\sc ii} systems in QSO 
spectra was poorly known because their number density 	
rapidly decreases with increasing $w_{\rm r}$(Mg\,{\sc ii}) (e.g. Steidel 	
\& Sargent \cite{steidel}), and only the SDSS data releases provided a robust 
measurement to be used for the purpose of comparison with the GRB results. 
As time elapsed, the number of afterglows observed at a resolution and 
signal-to-noise ratio (S/N)
sufficient to detect weaker systems increased, allowing Vergani et al. 
(\cite{vergani}) to measure the incidence of Mg\,{\sc ii} systems with
$0.3 < w_{\rm r}$(Mg\,{\sc ii}2796)$ <1.0$ ~\AA\ and to show that the latter do 
not display an excess towards GRBs as compared to QSOs. Thus the GRB excess is 
only significant for strong Mg\,{\sc ii} systems. 
	
Several physical effects have been proposed to explain the GRB-QSO 	
absorber discrepancy: 1) obscuration by dust of background QSOs, 2) 	
intrinsic nature of the GRB absorbing gas, 3) gravitational lensing of  	
GRBs by Mg\,{\sc ii} absorbers, 4) absorbers with small cores of a similar 	
size as GRB and QSO beam sizes (Prochter et al. \cite{prochtera}; 
Frank et al. \cite{frank}; Hao et al. \cite{hao}). 

In their study, Prochter et al.  (\cite{prochtera}) mention a possible 
excess of the same kind for strong Mg\,{\sc ii} systems towards BL Lac 
objects, as inferred by Stocke \& Rector (\cite{stocke}). BL Lac objects  
can be of great interest in our context because these targets are mainly 	
radio-selected and thus should not suffer from any extinction bias.	
Furthermore, the timescale and amplitude of their optical variations is 	
intermediate as compared to those of QSOs and GRBs. Finally, relativistic 
boosting is thought to be present in BL Lac objects, as in GRBs.  	
	
Since the result obtained by Stocke \& Rector was based on relatively few 	
 BL Lac spectra, not all with good S/N,  and thus of poor statistical	
significance, we have undertaken an observational programme devoted to the 	
measurement of $dN/dz$ towards Blazars for strong and weak Mg\,{\sc ii} systems. 
Indeed BL Lacs belong to the broader Blazar class characterized by violent variability, 
a compact flat-spectrum radio source, and a smooth optical-near infrared continuum. 	
Since the available data on absorption systems in BL Lac objects with the 
appropriate resolution and S/N has been limited so far, our survey 
can also provide valuable results on systems with an	
intrinsic nature, to be compared with the detailed information provided by	
the SDSS data on associated systems in QSO spectra. One might expect	
for instance differences related to the nature of the host galaxies 	
or to the presence of interstellar gas entrained by the jet.	
	
The paper is organised as follows. In Sect. 2, we describe how targets 
were selected. We present the 
observations and raw results in Sect. 3. Results on the incidence of 
intervening and associated Mg\,{\sc ii} systems  are given in Sect. 4. 
In Sect. 5, we compare our results with those obtained for GRBs and QSOs
and investigate possible interpretations of the excess of Mg\,{\sc ii} Blazar 
absorption systems. We give our summary  in Sect. 6.

%
\begin{table}  	
\caption[]{The sample.}	
\begin{center}
\begin{tabular}{l@{\hspace{1.4mm}}c@{\hspace{1.5mm}}c@{\hspace{1.6mm}}c@{\hspace{1.5mm}}c@{\hspace{1.4mm}}r}
\hline	
\noalign{\smallskip}	
  target & class & coordinates& $z_{\rm em}$ & grism & $\Delta t/n$ \\	
common name  & & 2000 & & & min \ \ \ \\	
\noalign{\smallskip}	
\hline	
\noalign{\smallskip}	
PKS 0057$-$338  & BL &  010009.0$-$333730  &  0.875  & B600 & 120/3 \\	
PKS 0208$-$512 $^{\gamma}$  & BL &  021046.2$-$510101  & 1.003 &  B600  & 80/2  \\	
PKS 0215$+$015$^{PG}$$^{\gamma}$$^C$  & BL &  021748.9$+$014449  & 1.715 & V600 & 60/2  \\ 	
2QZ  & opt &  023405.5$-$301519 & 1.690 & V600 & 80/2   \\	
PKS 0235$+$164$^{RS}$$^{\gamma}$$^C$  & BL &  023838.9$+$163659  & 0.940 & B600 & 60/3   \\	
SDSS       & opt &  024156.4$+$004351  & 0.989 & B600 & 120/3  \\	
PKS 0256$+$075  & BL&  025927.0$+$074739  & 0.893 & B600 & 80/2   \\	
PKS 0332$-$403 $^{\gamma}$  & BL &  033413.6$-$400825  & \ \ 1.351$^{N}$ & V600  & 115/3 \\	
PKS 0420$-$014 $^{\gamma}$$^C$  & FS &  042315.8$-$012033  & 0.915 & B600 & 80/3   \\	
PKS 0426$-$380$^{RS}$$^{\gamma}$  & BL &  042840.4$-$375619  & 1.110 & B600 & 80/2    \\	
PKS 0454$-$234 $^{\gamma}$$^C$  & BL &  045703.1$-$232452  & 1.003 & B600 & 90/3   \\	
PKS 0537$-$441$^{RS}$$^{\gamma}$$^C$  & BL &  053850.3$-$440508  & 0.890 & B600  & 20/4  \\	
PKS 0743$-$006  & FS &  074554.0$-$004417  & 0.994 & B600 & 100/3   \\	
PKS 0808$+$019$^{PG}$$^C$  & BL &  081126.7$+$014652  & 1.145 & B600 & 80/2   \\	
PKS 0823$-$223$^{PG}$$^{\gamma}$  & BL &  082601.5$-$223027  &  $\geq$0.910$^{\rm a}$ & B600 & 20/2\\
PKS 0906$+$015  & FS &  090910.1$+$012135  & 1.022 & B600 & 25/1   \\	
SDSS       & BL &  094257.8$-$004705  & 1.362 & V600 & 80/2   \\	
           & & & & B600 & 40/1 \\	
SDSS$^{Pl}$       & BL &  094827.0$+$083940  & 1.489 & V600 & 80/2   \\	
SDSS       & BL &  100959.6$+$014533  &  $\geq$1.085$^{\rm a}$ & V600 & 120/3  \\	
PKS 1008$+$013  & BL &  101115.6$+$010642  & \ \ 1.275$^{N}$ & V600 & 80/2   \\	
PKS 1130$+$008  & BL &  113245.6$+$003427  & 1.234 & V600 & 60/2   \\	
PKS 1144$-$379$^{RS}$$^{\gamma}$  & BL &  114701.3$-$381211  & 1.049 & B600 & 40/2   \\	
SDSS       & BL &  124533.7$+$022825  & $\geq$1.096$^{\rm a}$ & B600 & 80/2  \\	
           & & & & V600 & 80/2 \\	
PKS 1250$-$330$^{PG}$  & BL &  125258.3$-$331959   & 0.856 & B600 & 120/3  \\	
FIRST$^{Pl}$      & BL &  133859.0$+$115316  & $\geq$1.589$^{\rm b}$ & V600 & 80/2  \\	
PKS 1406$-$076 $^{\gamma}$$^C$  & FS &  140856.4$-$075226  & 1.500 & V600 & 80/2 \\
PKS 1407$+$022$^{PG}$  & BL &  141004.6$+$020306  & \ \ 1.253$^{N}$ & V600 & 80/2  \\	
SDSS       & BL &  141927.4$+$044513  & $\geq$1.684$^{\rm a}$ & V600 & 80/2    \\	
           & & & & R600 & 30/1 \\	
PKS 1424$-$418 $^{\gamma}$  & FS &  142756.2$-$420619  & 1.522 & V600 & 57/2  \\	
PKS 1519$-$273$^{RS}$  & BL &  152237.6$-$273010  & 1.294 & V600 & 115/3  \\	
SDSS       & BL &  154515.7$+$003235  & 1.051 & V600 & 80/2 \\	
PKS 1741$-$038  $^{\gamma}$$^C$  & FS &  174358.8$-$035004  & 1.054 & B600 & 80/2    \\
PKS 1953$-$325  & FS &  195659.4$-$322546 & 1.242 & V600 & 120/3 \\	
PKS 2012$-$017$^{PG}$  & BL &  201515.1$-$013732  & $\geq$0.940$^{\rm c}$  & B600 & 80/2  \\
PKS 2029$+$121$^{RS}$  & BL &  203155.0$+$121941  & 1.215 & V600 & 80/2   \\	
PKS 2131$-$021$^{RS}$$^C$  & BL &  213410.3$-$015317  & 1.285 & V600 & 80/2   \\	
MH  2133$-$449$^{PG}$  & BL &  213618.3$-$444348  & $\geq$0.980$^{\rm c}$ & B600 & 120/3 \\
EQ 2207.3$+$004$^{PG}$ & BL &  220719.7$+$004157 & 1.892 & V600 & 80/2 \\	
2QZ  & opt &  221450.1$-$293225 & 1.636 & V600 & 80/2  \\	
PKS 2223$-$114$^{PG}$  & BL &  222543.7$-$111341   & 0.997 & V600 & 80/2  \\ 
PKS 2223$-$052$^C$  & BL &  222547.2$-$045701  & 1.404 & V600 & 80/2   \\	
PKS 2308$-$109  & BL &  231116.9$-$103849  & 1.529 & V600 & 80/2   \\
\noalign{\smallskip}	
\hline		
\noalign{\smallskip}
B2 0218$+$35$^{RS}$$^{\gamma}$   & BL &  022105.4$+$355615 &  0.944 & N/A  & N/A  \\
S5 0454$+$844$^{RS}$  & BL &  050842.4$+$843204 &  $\geq$1.340 & N/A  & N/A  \\
B2 1308$+$32$^{RS}$$^{\gamma}$$^C$   & BL &  131028.7$+$322044 &  0.997 & N/A  & N/A  \\ 
\noalign{\smallskip}	
\hline	
\noalign{\smallskip}
\multicolumn{6}{l}{class - BL: BL Lac object, opt: optical BL Lac candidate, FS: Flat }\\
\multicolumn{6}{l}{\hspace{3mm} Spectrum Radio QSO (FSRQ).}\\
\multicolumn{6}{l}{$n$ \ Number of exposures.}\\
\multicolumn{6}{l}{$^{RS}$ In the 1 Jy sample of Rector \& Stocke (2001).}\\
\multicolumn{6}{l}{$^{PG}$ In the BL Lac sample of Padovani \& Giommi (1995) and not in RS.}\\
\multicolumn{6}{l}{$^{Pl}$ In the BL Lac list kindly provided by R. Plotkin prior 
to publication.}\\
\multicolumn{6}{l}{$^{\gamma}$ Gamma-ray emitter.}\\
\multicolumn{6}{l}{$^C$ Superluminal motions.}\\
\multicolumn{6}{l}{$^{N}$ New emission redshift.}\\
\multicolumn{6}{l}{$^{\rm a}$ Upper limit derived from Mg\,{\sc ii} absorption.}\\
\multicolumn{6}{l}{$^{\rm b}$ Upper limit derived from C\,{\sc iv} absorption.}\\
\multicolumn{6}{l}{$^{\rm c}$ Upper limit derived from lack of Ca\,{\sc ii} absorption
from the host} \\
\multicolumn{6}{l}{\hspace{3mm} galaxy.}\\
\end{tabular}	
\end{center}	
\label{obs}	
\end{table}	
%
%
	
\section{The sample}	
%

We aim at measuring $dN/dz$ for Mg\,{\sc ii} intervening absorption systems 
towards BL Lac objects  with both strong and weak rest equivalent width, 
in order to check whether the excess, if real, depends on the strength of 
the Mg\,{\sc ii} absorption doublet. This is not an easy task for several reasons. 
First, our project requires targets with known emission redshifts to 	
properly estimate the redshift interval probed. Second,  BL Lacs  	
and more generally Blazars exhibit large, unpredictible magnitude fluctuations,  
making it difficult to obtain a homogeneous data set. As in previous	
studies, we shall consider for our statistical analyses that systems with a 
relative velocity $\Delta v > 5000$ km s$^{-1}$ from the source are of 
intervening nature while the others are considered as associated.

\subsection{Target selection }
\label{target}	
	
Since, at low redshift ($z<1.2$), there is a strong  cosmic evolution of the 	
number density of QSO  Mg\,{\sc ii} absorbers with $w_{\rm r}$(Mg\,{\sc ii}2796) 	
$>1$~\AA\ (Nestor et al. \cite{nestor}; Prochter et al. \cite{prochterb}),  
we selected targets with 
an emission redshift $z_{\rm em}>0.8$ to optimize the efficiency of our 	
spectroscopic observations. This also yields a redshift range comparable to 
that investigated for GRBs by Prochter et al. (\cite{prochtera}). 
We restricted further the redshift range to  $z_{\rm em} \lesssim1.8$ due to  
the paucity and faintness of Blazars at higher redshift.

Our sample comprises the seven  1~Jy BL Lacs of Rector \& Stocke (\cite{rector}) 
at $z_{\rm em}>0.8$ and DEC $< +20 \deg$ to: 1) confirm the strength of the 
Mg\,{\sc ii} absorption lines for spectra of low S/N, 2) detect weaker 
Mg\,{\sc ii}, 3) complement the available information on emission redshifts.  
The same selection criteria were applied to the BL Lac sample of Padovani 
\& Giommi (\cite{padovani95}) (nine additional targets).  
We have taken into account recent emission redshift measurements, including 
upper limits (Sbarufatti et al. \cite{sbarufatti5} \& \cite{sbarufatti6}). 
The latter are derived from either the Mg\,{\sc ii} absorption system of  
highest redshift or the absence of the Ca\,{\sc ii} absorption doublet from 
the host galaxy in the BL Lac spectrum. We have re-observed almost all the 
$0.9<z_{\rm em}<1.8$  BL Lacs studied by these authors, since 
the  Mg\,{\sc ii} absorption doublets are not resolved in their  low resolution 
FORS1 data (FWHM = 15-20 \AA\ with the 300V+I grism). 	

To increase our sample, we searched for targets, at $0.8<z_{\rm em}<1.8$ 
and DEC $< +20 \deg$, of the  Blazar class in the SIMBAD and NED databases. 
In addition to BL Lacs, the Blazar class also includes Flat Spectrum Radio QSOs 
(FSRQs) which exhibit broad emission signatures (Angel \& Stockman \cite{angel}). 
Blazars are thought to have jets pointing towards us with associated emission beamed 
by relativistic motions. When in high state, the optical emission can be highly 
polarized (e.g. Impey \& Tapia \cite{impey8} \& \cite{impey9}).  	
When in low state, the isotropic thermal emission from the active	
nucleus and its associated broad emission line region can be detectable or 	
even dominate over the Doppler boosted emission from the jet. 
The magnitudes given in SIMBAD only reflect the state of the object at 
a given epoch; nevertheless, we introduced a V magnitude cut of 21.0 to 
exclude (not always successfully) very faint sources.	

This SIMBAD and NED searches provided most of the remaining targets of our sample. 
It includes optically selected BL Lacs among the 2DF AAT  QSO redshift survey  and 
the SDSS (Londish et al. \cite{londish} \& \cite{londish7}; Collinge et al. \cite{collinge}). 
Some of the SDSS BL Lacs have high optical polarization which confirms the 
synchroton nature of the optical emission (Smith  et al. \cite{smith}). 


Finally,  we considered information kindly provided by Plotkin on seven 
'southern' high-redshift BL Lacs prior to publication (Plotkin et al. \cite{plotkin})
of their BL Lac sample selected from a combination of the SDSS Data Release 5 (DR5) 
and the VLA FIRST survey. Only two of them, at $z_{\rm em} <1.8$, were not already in 
our SIMBAD/NED list (see Table \ref{obs}); both were observed. The upper limit on 
$z_{\rm em}$ for one of them is derived from a C\,{\sc iv} absorption doublet 
detected in the SDSS spectrum. 
	
The resulting sample comprises 51 sources for our observations with FORS1
scheduled in November 2007 and May 2008. Due to crowding in right ascension 
(RA) around 11- 12h, we could only observe 42 of them; these were 
chosen	according to their RA and DEC only, disregarding any previously	
available information on absorption systems. 
To consistently take into account all the 1~Jy sources of Rector \& Stocke 
(\cite{rector})  at $z_{\rm em}>0.8$, we included the three northern sources
which have published, reliable spectroscopic information on the presence or 
lack of strong Mg\,{\sc ii} systems (Miller et al. 
\cite{miller}; Stocke \& Rector \cite{stocke}; Cohen et al. \cite{cohen}). 

Thus, there are 45 sources at $0.8 <z_{\rm em} \lesssim1.8$ in our final sample
of which ten BL Lacs from the 1~Jy sources of Rector \& Stocke (\cite{rector}). 
 Our sample comprises  35 BL Lacs, 7 FSRQs and 3 optical BL Lac 
candidates. Among the radio-loud Blazars, there are 
15 gamma-ray sources, this emission being an important component of the 
relaticistic jet (Mukherjee et al. \cite{mukherjee}; Abdo et al. \cite{abdo}). 
Superluminal motions are detected in 11 radio Blazars 
(Piner et al. \cite{piner}, \cite{piner07}; Fan et al. \cite{fan09}) 
and, of those, eight are indeed also known gamma-ray emitters. 
Our 42 observed sources are presented in Table \ref{obs} as well as 
information on the three northern BL Lacs. 	
We now discuss possible biases affecting this sample.

	
\subsection{Potential biases }	
%

In order to obtain a reliable measurement of $dN/dz$, one must carefully 	
identify potential selection effects that might bias the estimate	
towards larger (or smaller) values. 

One first difficulty is related to	
the fact that for some sources, no emission lines had been detected or	
identified at the time of our sample selection and only a lower limit on	
$z_{\rm em}$ was provided by a Mg\,{\sc ii} system. Including such systems in	
the statistics would clearly bias $dN/dz$ towards a too high value 
since they might in fact be associated instead of intervening. 
To avoid this bias, we discard these possibly associated Mg\,{\sc ii} systems,  
if our observations do not reveal new emission lines providing a firm 
source redshift. For these cases, we  include in the analysis only the lower 
redshift part of the sightline (i.e. beyond 5000 km s$^{-1}$ of the absorption 
redshift).  This is a conservative approach, since some of the 'possibly 
associated' Mg\,{\sc ii} systems could be intervening.  
Four of our targets have $z_{\rm em}$ constrained by a strong  Mg\,{\sc ii} 
absorber, and one by a high redshift C\,{\sc iv} absorber without associated 
Mg\,{\sc ii}. On the opposite, if our spectrum provides a new	
firm redshift for the source, all intervening Mg\,{\sc ii} systems are included.
	
A second concern involves sources for which intervening systems were already 
known to be present. Ideally, one should observe for our programme only sources 
for which no information at all was available on the presence or absence of 
absorption systems. Unfortunately, this would severely restrict the number 
of targets, which is already small. To 
circumvent this difficulty, we performed our selection exclusively on	
the basis of emission properties. If those targets with already known	
systems were excluded, this would bias $dN/dz$ towards {\it lower} values. 	
Although spectra exist for sources like PKS 0235$+$164 and PKS 0215$+$015, 	
we nevertheless reobserved them to obtain either a more complete 	
wavelength coverage or  a uniform $w_{\rm r}$(Mg\,{\sc ii}2796) limit. 
 Further, this offers the opportunity to investigate temporal variations in 
these systems over a reasonably large time interval, since some of them were 
initially detected at high S/N about 25 years ago (for PKS 0215$+$015 see Bergeron 
\& D'Odorico \cite{bergeron86d}; we will present our results on possible 
variations of Mg\,{\sc ii} and Fe\,{\sc ii} absorptions in a forthcoming paper).

Finally, given the broad property range of the Blazar class, 
it remains possible that some targets be in fact QSOs. 
Since our main purpose is to test 
whether the number density of strong Mg\,{\sc ii} systems is {\it larger} 
towards Blazars, this is however not a real problem because contamination 
of our sample by QSOs would then {\it lower} our $dN/dz$ estimate. 
		
	
\section{Observations }	
%

%
\begin{table*}   	
\caption[]{Mg\,{\sc ii} absorption systems.}	
\begin{center}	
\begin{tabular}{cclcccl}	
\hline	
\noalign{\smallskip}	
  target & $z_{\rm em}$ &  emission lines & $z_{\rm abs}$ &$ w_{\rm r}$(2796,2803)$^u$ & 	
$w_{\rm r}$(2586,2600)$^u$ & associated absorptions  \\	
  & & & & \AA\ \ \  \AA\ &\AA\ \ \  \AA\  &  \\	
\noalign{\smallskip}	
\hline	
\noalign{\smallskip}	
 0100$-$3337  & 0.875 & C\,{\sc iii}],Mg\,{\sc ii} & 0.6810 & 0.36,0.36  & nd,nd  & -- \\	
 0217$+$0144  & 1.715  & C\,{\sc iii}]  & 0.7557 & 0.12,0.06  & nd,nd  & --  \\ 	
  &   &  & 0.9685 & 0.22,0.15  & nd,nd & --   \\ 	
  &   &  & 1.3439 &  {\bf 1.92,1.83} & 0.75,1.39 &  Mg{\tt I},Fe\,{\sc ii},	
Mn\,{\sc ii},Zn\,{\sc ii},Cr\,{\sc ii}    \\
 0234$-$3015  & 1.690 & Al\,{\sc iii}],C\,{\sc iii}] &  1.2361 & 0.11,0.09  & nd,nd & --     \\		
 0238$+$1636  & 0.940 &  C\,{\sc iii}],Mg\,{\sc ii} & \ \ \ 0.5245$^{SR}$ & {\bf 2.26,2.12} 
& 0.89,1.63 &	Mg{\tt I},Fe\,{\sc ii},Mn\,{\sc ii}    \\	
   &  &   & \ \ \ 0.8522$^{SR}$  & 0.43,0.23 & nd,0.09 & Fe\,{\sc ii}    \\	
 0241$+$0043  & 0.989 & C\,{\sc iii}],Mg\,{\sc ii} & 0.6310 & 0.27,0.18 & nd,nd & -- \\
   &  &   & 0.7753 & {\bf 1.06,0.81} & 0.21,0.39 &  Fe\,{\sc ii}    \\	
   &  &   & \ \  0.9790$^{A}$ & 0.64,0.38 & nd,nd & Al\,{\sc iii}   \\
 0334$-$4008  & \ \  1.351$^N$ & Mg\,{\sc ii} & 1.0791 & 0.20,0.13 & nd,nd  & --    \\	
  &    &  & 1.2083 & 0.84,0.77 & 0.16,0.31  & Fe\,{\sc ii}   \\	
 0423$-$0120  & 0.915 & C\,{\sc iii}],Mg\,{\sc ii} & 0.6338  & 0.88,0.70 & 0.05,0.14 
& Mg{\tt I}, Fe\,{\sc ii}  \\	
 0428$-$3756  & 1.110 & Mg\,{\sc ii} & 0.5592 & 0.93,0.72 & 0.19,0.42 & Mg{\tt I},Fe\,{\sc ii}   \\	
  &  &   & \ \ \ 1.0283$^{SR}$  & 0.56,0.41 & nd,0.08 & Fe\,{\sc ii},Al\,{\sc iii}    \\	
 0457$-$2324  & 1.003 & Mg\,{\sc ii} & 0.8922 & {\bf 2.20,1.90} & 1.38,1,87 &  Mg{\tt I},Fe\,{\sc ii},	
Mn\,{\sc ii},Zn\,{\sc ii},Cr\,{\sc ii},Al\,{\sc iii}   \\	
 0538$-$4405  & 0.890 & Mg\,{\sc ii} & 0.6725 & \ \ 0.065,0.041$^a$ & nd,nd & --    \\	
 0745$-$0044  & 0.994 & C\,{\sc iii}],Mg\,{\sc ii} & 0.7979  & 0.095,0.056 & nd,nd & --    \\	
 0826$-$2230  & $\geq$0.910 \ \ & --  & 0.7057  & 0.11,0.06 & nd,nd & --    \\ 	
 &  &   & \ \ \ 0.9107$^{A?}$  & {\bf 1.28,0.92}  & 0.19,0.41 & Mg{\tt I},Fe\,{\sc ii}    \\	
 0909$+$0121  & 1.022 & C\,{\sc iii}],Mg\,{\sc ii} & 0.5369 & 0.41,0.31 & nd,0.14 & Fe\,{\sc ii}  \\	
 0942$-$0047  & 1.362 & Mg\,{\sc ii},broad abs  & 0.8182 & {\bf 1.58,1.28} & 0.70,0.96 & 
Mg{\tt I},Fe\,{\sc ii},Mn\,{\sc ii},Zn\,{\sc ii},Cr\,{\sc ii}  \\
  &   &  &  1.0231 & 0.38,0.22 & \ \ \ nd,nd & --   \\	
 0948$+$0839  & 1.489 & C\,{\sc iii}],Mg\,{\sc ii} & 1.0763  & 0.76,0.63 & \ \ \ nd,0.26 
& Fe\,{\sc ii}\\	
  &  &   & 1.3273 & {\bf 4.07,3.72} & 1.80,2.80 & Mg{\tt I},Fe\,{\sc ii},Mn\,{\sc ii}    \\
  &  &   & 1.4247 & {\bf 1.08,0.62} & \ \ 0.16,0.37: \  & Fe\,{\sc ii}    \\	
 1009$+$0145  & $\geq$1.085 \ \ & -- & \ \ \ 1.0851$^{\rm A?}$ & {\bf 1.23,0.90} & 0.18,0.32 & Fe\,{\sc ii}  \\	
  1147$-$3812 & 1.049 & C\,{\sc iii}],C\,{\sc ii}],Mg\,{\sc ii}  & 0.3750 & 0.27,0.20 & -- & --  \\
 1245$+$0228  & $\geq$1.096 \ \ & -- & 1.0116 & 0.60,0.48 & 0.26,0.44 & Fe\,{\sc ii}  \\	
  &  & & \ \ \ 1.0946$^{A?}$ & {\bf 2.30,2.27} & 1.82,2.04 & Mg{\tt I},Fe\,{\sc ii},Mn\,{\sc ii},Zn\,{\sc ii},
Cr\,{\sc ii},Al\,{\sc ii},Al\,{\sc iii},Si\,{\sc ii}  \\
 1408$-$0752  & 1.500 & C\,{\sc iii}],C\,{\sc ii}],Mg\,{\sc ii} & 1.2753 & {\bf 2.08,1.90} & 1.17,1.53 & 	
Mg{\tt I},Fe\,{\sc ii},Mn\,{\sc ii},Zn\,{\sc ii},Cr\,{\sc ii}    \\	
  &  &   & 1.2913 & 0.27,0.26 & \ \ \ nd,0.06: & Fe\,{\sc ii}   \\	
 1410$+$0203  & \ \ 1.253$^N$ & C\,{\sc ii}],Mg\,{\sc ii}  & 1.1123 & 0.79,0.56 & \ \ \ nd,0.08:
& Fe\,{\sc ii}   \\	
 1419$+$0445  & $\geq$1.684 \ \ & --  & 0.8987 & \ \ 0.73,0.91$^b$ & 0.21,0.35 & Fe\,{\sc ii}  \\
  &  &   & 1.0849 & 0.47,0.37 & 0.13, $b$ & Fe\,{\sc ii}   \\
  &  &   & 1.1039 & 0.52,0.25 & nd,nd & -- \\
  &  &   & 1.1834 & 0.12,0.09 & nd,nd & -- \\
  &  &   & 1.2722 & 0.62,0.58 & $b$ ,0.49 & Fe\,{\sc ii}   \\	
  &  &   & \ \ \ 1.6832$^{A?}$ & {\bf 2.60,1.89} & 0.49,0.80 & Fe\,{\sc ii}    \\	
 1427$-$4206  & 1.522 & C\,{\sc iii}],C\,{\sc ii}],Mg\,{\sc ii} & 1.0432  & \ 0.23,0.12: & nd,nd & -- \\	
  &  &  &  1.0907  & 0.62,0.37 & nd,0.09 & Fe\,{\sc ii}    \\
 1522$-$2730 & 1.294 & Mg\,{\sc ii} & \ \ 1.2847$^{A}$ & 0.25,0.22  & nd,0.06 & Fe\,{\sc ii}  \\	
 1743$-$0350  & 1.054 & C\,{\sc iii}],C\,{\sc ii}],Mg\,{\sc ii} & 0.2527 & \ 0.64,0.49: & -- & \\
  &  &  & 0.5293 & {\bf 1.05,0.80} & \ \ 0.37,0.65$^b$ & Mg{\tt I},Fe\,{\sc ii}   \\	
  &  &  & 0.9077 & {\bf 2.46,2.05} & 0.89,1.32 & Mg{\tt I},Fe\,{\sc ii},Mn\,{\sc ii}  \\
 1956$-$3225  & 1.242 & Mg\,{\sc ii} & 0.6236 & 0.95,0.90 & -- & -- \\
 & & & 1.0660 & {\bf 3.53,3.49} & 1.77,2.70 &	
Mg{\tt I}$^b$,Fe\,{\sc ii},Mn\,{\sc ii}   \\	
 2031$+$1219  & 1.215 &  Mg\,{\sc ii}  & \ \ \ 1.1158$^{SR}$  &  {\bf 1.29,1.16} & 0.71,1.12 &  	
Mg{\tt I},Fe\,{\sc ii},Mn\,{\sc ii}   \\
  2134$-$0153 & 1.285 & Mg\,{\sc ii} & \ \ 1.2458$^{A}$ & 0.22,0.15 & nd,nd &  --  \\	
 2136$-$4443  & $\geq$0.980 \ \  & -- & 0.5211 & 0.51,0.39 & \ 0.12:,0.26 & Mg{\tt I},Fe\,{\sc ii}   \\
 2214$-$2932  & 1.636 & Al\,{\sc iii},C\,{\sc iii}] &   1.5070 & 0.38,0.33 & nd,0.17 &  Fe\,{\sc ii}  \\		
 2225$-$0457  & 1.404 & C\,{\sc iii}],C\,{\sc ii}],[Ne\,{\sc iv}],Mg\,{\sc ii}  &  0.8458  & 0.54,0.43 & 0.05,0.18 
& Fe\,{\sc ii}    \\	
 &    &  & 0.9782 & 0.10,0.07  & nd,nd & --     \\	
 2311$-$1038  & 1.529 & C\,{\sc iii}],C\,{\sc ii}] &  \ \ 1.5242$^{A}$  &  {\bf 3.61,3.09} & 1.37,2.34 & 	
Mg{\tt I},Fe\,{\sc ii},Al\,{\sc iii}  \\	
\noalign{\smallskip}  	
\hline	
\noalign{\smallskip}
0221$+$3556 & 0.944 & Mg\,{\sc ii} & 0.6850$^{SR}$ &  {\bf 1.78,1.72 } & na,na  &  Fe\,{\sc ii},Mg{\tt I}  \\
\hline	
\noalign{\smallskip}
\multicolumn{7}{l}{$^u$ Representative uncertainties are given in Sect. 3.2.}\\
\multicolumn{7}{l}{$^{N}$ New emission redshift.}\\
\multicolumn{7}{l}{$^{SR}$ Strong Mg\,{\sc ii} system in Stocke \& Rector (1997).}\\	
\multicolumn{7}{l}{$^A$ Associated system.}\\	
\multicolumn{7}{l}{$^{A?}$ Possibly associated system.}\\ 
\multicolumn{7}{l}{$^a$ Target in very bright state. The optical flux is known to vary by 
a factor of at least 35.}\\	
\multicolumn{7}{l}{$^b$ Blended with an other absorption.}\\	
\multicolumn{7}{l}{nd : not detected.}\\	
\multicolumn{7}{l}{na : not available.}	
\end{tabular}	
\end{center}	
\label{MgP8081}	
\end{table*}	

\subsection{Observations and data reduction}	

The observations were performed  with the FOcal Reducer and low-dispersion 
Spectrograph (FORS1) at the ESO VLT UT2 (Kueyen) telescope in November 2007 
and May 2008 (Visitor Mode). During these observing runs, the seeing was in 
the range 0.5$\arcsec$ to 1.3$\arcsec$, with an average of 0.8$\arcsec$. 
In the long-slit mode, the selected B600 and V600 grisms covered the 
3405-6095~\AA\  and 4435-7390~\AA\ spectral ranges, respectively, and yielded a 
spectral resolution, measured on our target spectra, of  FWHM$=5.0\pm0.5$~\AA\ (or 
$3.4\pm0.34$ binned pixels) for our 1.0$\arcsec$ slit width. Details on the exposure 
times and grisms are given in Table \ref{obs}. 
Data reduction was performed using MIDAS (08FEBpl1.1). The double exposures for each 
target enabled an adequate correction of cosmic ray contamination. Standard spectroscopic 
stars (HILT 600, LTT 9239 and for both observing runs LTT 3218) were observed with 
a 5.0$\arcsec$ slit width. The seeing although general good was nevertheless fairly 
variable and thus only a relative  flux calibration could be performed.

   \begin{figure}	
   \centering	
      \includegraphics[width=8cm]{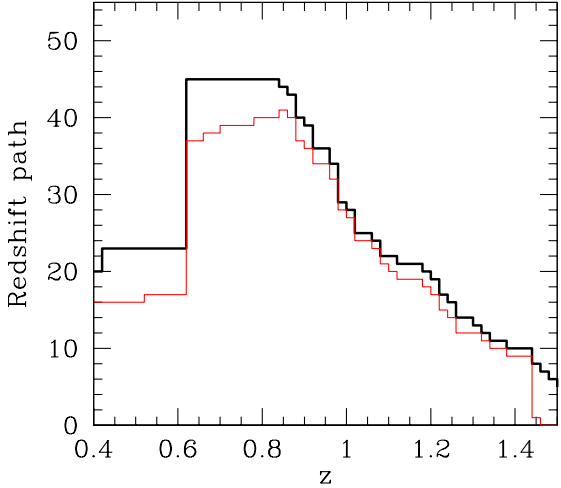}	
    \caption{Redshift path density for the intervening 
   Mg\,{\sc ii} systems towards Blazars: strong 	
  (w$_r (2796) > 1.0$~\AA: solid black curve) and weak ones	
  ($0.3<$ w$_r (2796) < 1.0$~\AA: thinner red curve).}	
         \label{zpath}	
   \end{figure}	

\subsection{ Redshift and equivalent width estimates}	
	
Among the 42 Blazars observed with FORS1, we detect Mg\,{\sc ii} and/or C\,{\sc iii}] 
or C\,{\sc ii}] emission lines in 35 sources. All the detected emission lines are 
given in Table \ref{MgP8081}. Some of our $z_{\rm em}$ estimates differ from those 
those already published, including SDSS ones, as our spectra are of higher S/N. 
We could determine new emission redshift (labelled $N$ in Tables \ref{obs} and 
\ref{MgP8081}) for three Blazars, of which one without any Mg\,{\sc ii} absorption 
system. Their spectra are presented in Appendix A.  	
	

Mg\,{\sc ii} absorption doublets are detected in 31 Blazars observed with FORS1 
plus in one of the three northern 1~Jy BL Lac. Their absorption redshift, $z_{\rm abs}$, 
and rest equivalent widths are listed in Table \ref{MgP8081}. We also give the rest 
equivalent widths of their associated Fe\,{\sc ii}2586,2600 doublets together with a 
list of all the other detected absorptions. 
 Uncertainties on $w_{\rm obs}$ values vary depending on the spectral region and target 
magnitude. Representative values are in the range 0.04 - 0.12 \AA\ (corresponding to 
relative uncertainties of about 2 to 10\% depending on line strength) for intermediate 
S/N values of 50 to 100. In our estimate, we include both pixel-to-pixel noise and uncertainty
in the continuum placement; the latter is dominant when the S/N is larger than about 100. 
In some cases, the Blazars were in a very high state with an optical flux fully dominated 
by synchrotron emission and we could then detect very weak Mg\,{\sc ii} absorption 
doublets down to $w_{\rm r}(2796) = 65$~m\AA. 

We do not confirm the large strength of the $z_{\rm abs} = 1.0283$ Mg\,{\sc ii} 
system in PKS 0426$-$380 from the sample of Stocke \& Rector (\cite{stocke}). 
The spectrum of this Blazar is presented in Appendix C.  

In one Blazar, SDSS J094257.8$-$004705, there is an unusual broad absorption line 
with a FWHM $\simeq 6000$ km s$^{-1}$ that we identify as a highly detached 
Mg\,{\sc ii} BAL at $z_{\rm abs} = 0.929$ or $\Delta v= 0.20$c. More information on 
this system is given in Appendix B. 
Known and potential damped Ly$\alpha$ systems are discussed in Appendix D.
	


\subsection{ Statistical sample definition}	

\begin{table}[t]    	
\caption[]{Redshift path for strong ($w_{\rm r}(2796)>1.0$~\AA) and weak
Mg\,{\sc ii} ($w_{\rm r}(2796)>0.3$~\AA) absorbers and 
S/N per pixel$^a$ around peak efficiency.}
\begin{center}	
\begin{tabular}{crccr}	
\hline	
\noalign{\smallskip}	
  target & $z_{\rm em}$ & $z$ path  & $z$ path & S/N \\	
 &  & Strong  & weak &  \\	
\noalign{\smallskip}	
\hline	
\noalign{\smallskip}	
 0100$-$3337  &  0.875 &  0.494 &  0.306 & 50  \\	
 0210$-$5101  &  1.003 &  0.620 &  0.620 & 95 \\	
 0217$+$0144  &  1.715 &  0.956 &  0.807 & 190 \\ 
 0234$-$3015  &  1.690 &  0.956 &  0.807 & 120 \\	
 0238$+$1636  &  0.940 &  0.558 &  0.558 & 80 \\	
 0241$+$0043  &  0.989 &  0.607 &  0.607 & 55 \\	
 0259$+$0747  &  0.893 &  0.511 &  0.145 & 40 \\	
 0334$-$4008  &  1.351 &  0.689 &  0.689 & 95 \\	
 0423$-$0120  &  0.915 &  0.533 &  0.533 & 180 \\	
 0428$-$3756  &  1.110 &  0.725 &  0.725 & 270 \\	
 0457$-$2324  &  1.003 &  0.620 &  0.620 & 350 \\	
 0538$-$4405  &  0.890 &  0.514 &  0.514 & 250 \\	
 0745$-$0044  &  0.994 &  0.611 &  0.611 & 200 \\	
 0811$+$0146  &  1.145 &  0.759 &  0.759 & 85 \\	
 0826$-$2230  &  $\geq$0.910 &  0.528 &  0.528 & 200 \\ 	
 0909$+$0121  &  1.022 &  0.634 &  0.634 & 140 \\	
 0942$-$0047  &  1.362 &  0.698 &  0.698 & 105 \\	
 0948$+$0839  &  1.489 &  0.812 &  0.812 & 55 \\	
 1009$+$0145  &  $\geq$1.085 &  0.427 &  0.191 & 40 \\	
 1011$+$0106  &  1.275 &  0.614 &  0.614 & 60 \\	
 1132$+$0034  &  1.234 &  0.563 &  0.563 & 130 \\	
 1147$-$3812  &  1.049 &  0.665 &  0.665 & 55 \\	
 1245$+$0228  &  $\geq$1.096 &  0.711 &  0.703 & 75 \\	
 1252$-$3319  &  0.856 &  0.475 &  0.000 & 28 \\	
 1338$+$1153  &  $\geq$1.589 &  0.923 &  0.807 & 150 \\	
 1408$-$0752  &  1.494 &  0.829 &  0.829 & 100 \\	
 1410$+$0203  &  1.256 &  0.595 &  0.595 & 75 \\	
 1419$+$0445  &  $\geq$  1.684 &  0.956 &  0.807 & 80 \\	
 1427$-$4206  &  1.522 &  0.857 &  0.807 & 100 \\	
 1522$-$2730  &  1.294 &  0.633 &  0.633 & 110 \\	
 1545$+$0032  &  1.051 &  0.394 &  0.394 & 50 \\	
 1743$-$0350  &  1.054 &  0.670 &  0.670 & 70 \\
 1956$-$3225  &  1.242 &  0.582 &  0.582 & 50 \\	
 2015$-$0137  &  $\geq$0.940  &  0.558 &  0.558 & 115 \\	
 2031$+$1219  &  1.215 &  0.555 &  0.498 & 55 \\	
 2134$-$0153  &  1.285 &  0.624 &  0.624 & 130 \\	
 2136$-$4443  &  $\geq$0.980 &  0.597 &  0.597 & 115 \\
 2207$+$0041  &  1.892 &  0.817 &  0.467 & 55 \\
 2214$-$2932  &  1.636 &  0.956 &  0.807 & 48 \\	
 2225$-$1113  &  0.997 &  0.614 &  0.176 & 35 \\ 	
 2225$-$0457  &  1.404 &  0.741 &  0.741 & 125 \\	
 2311$-$1038  &  1.529 &  0.864 &  0.807 & 75 \\	
\noalign{\smallskip}	
\hline	
\noalign{\smallskip}
 0221$+$3556  & 0.944  &  0.482 &  0.000 & N/A \\
 0508$+$8432  & $\geq$1.340  &  0.871 &  0.000 & N/A \\
 1310$+$3220  & 0.997  &  0.534 &  0.000 & N/A \\  
\noalign{\smallskip}	
\hline	
\multicolumn{4}{l}{$^a$	FWHM = 3.4 px.} \\
\end{tabular}	
\end{center}	
\label{S/N}
\end{table}

Our Blazar sample comprises 45 targets. Three are northern objects without 
FORS1 data, but included here as they are in the 1Jy sample discussed by 
Stocke \& Rector (\cite{stocke}). The S/N of their spectra only enables the 
detection of strong  ($w_{\rm r}(2796) > 1.0$~\AA) Mg\,{\sc ii} doublets. 
For the Blazars with unknown $z_{\rm em}$, the lower limit on $z_{\rm em}$ is 
set by the Mg\,{\sc ii} system of highest $z_{\rm abs}$. Associated Mg\,{\sc ii} 
absorbers, within 5000 km s$^{-1}$ of $z_{\rm em}$ (or its lower limit) are 
excluded from  our Mg\,{\sc ii} samples. 

The redshift path for all the sightlines is given in Table \ref{S/N} and shown in 
Fig. \ref{zpath}. 
 For both grisms, the minimum values $z_{\rm min}$ correspond 
to either the wavelengths at which the overall efficiency is half that at peak 
efficiency or, if larger, the redshift at which the S/N enables the 3$\sigma$ 
detection of the Mg\,{\sc ii} doublet (with $w_{\rm r}(2796)$ limits of either 
0.3 or 1.0 \AA). For the former, these values equal 
$z_{\rm min} = 0.350$ and 0.623 for the B600 and V600 grisms, respectively. 
The efficiency of the B600 grism strongly declines towards its blue end and,  
although one weak Mg\,{\sc ii} system at $z_{\rm abs} = 0.2527$ was detected 
in one sightline, the statistics are too poor to include this $z$ range in our study. 
Similarly, the S/N decreases at the red end of the V600 grism (see Fig. A.1) due 
to strong sky lines  which prevents detection of weak Mg\,{\sc ii} system at 
$z_{\rm abs} > 1.430$ for most sightlines. This value is adopted as the upper limit 
of our $z$ range for weak Mg\,{\sc ii} systems (the corresponding limit for strong 
systems is $z_{\rm abs} = 1.579$).  
Thus the  weak Mg\,{\sc ii} system at $z_{\rm abs} = 1.5070$ and the strong system 
at $z_{\rm abs} = 1.6832$ (special setting: R600 grism) given in Table \ref{MgP8081}
are excluded from our statistical samples.

\section{ Incidence of Mg\,{\sc ii} absorbers in Blazar spectra}

   \begin{figure*}	
   \centering	
      \includegraphics[width=6cm]{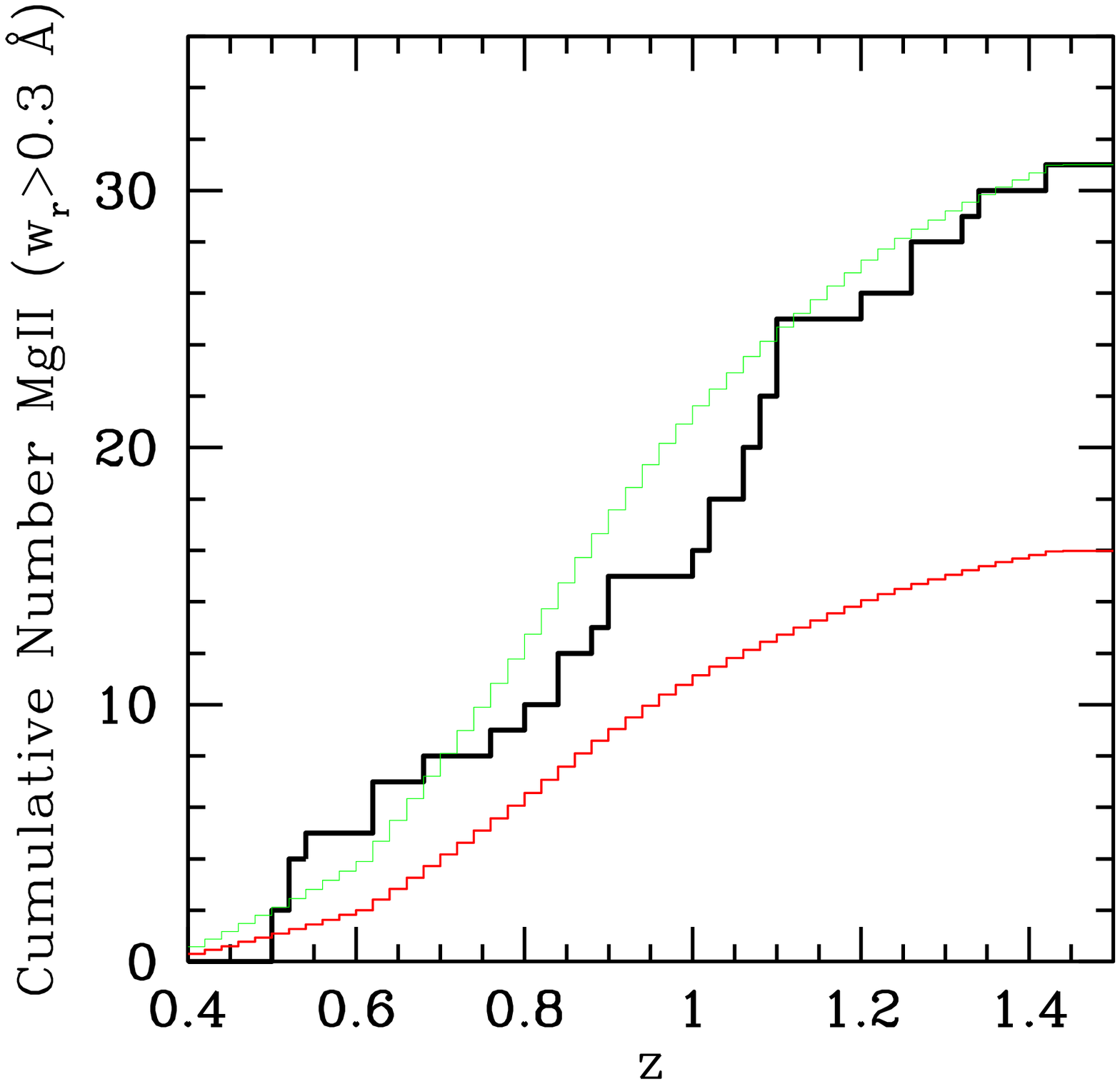}
      \includegraphics[width=6cm]{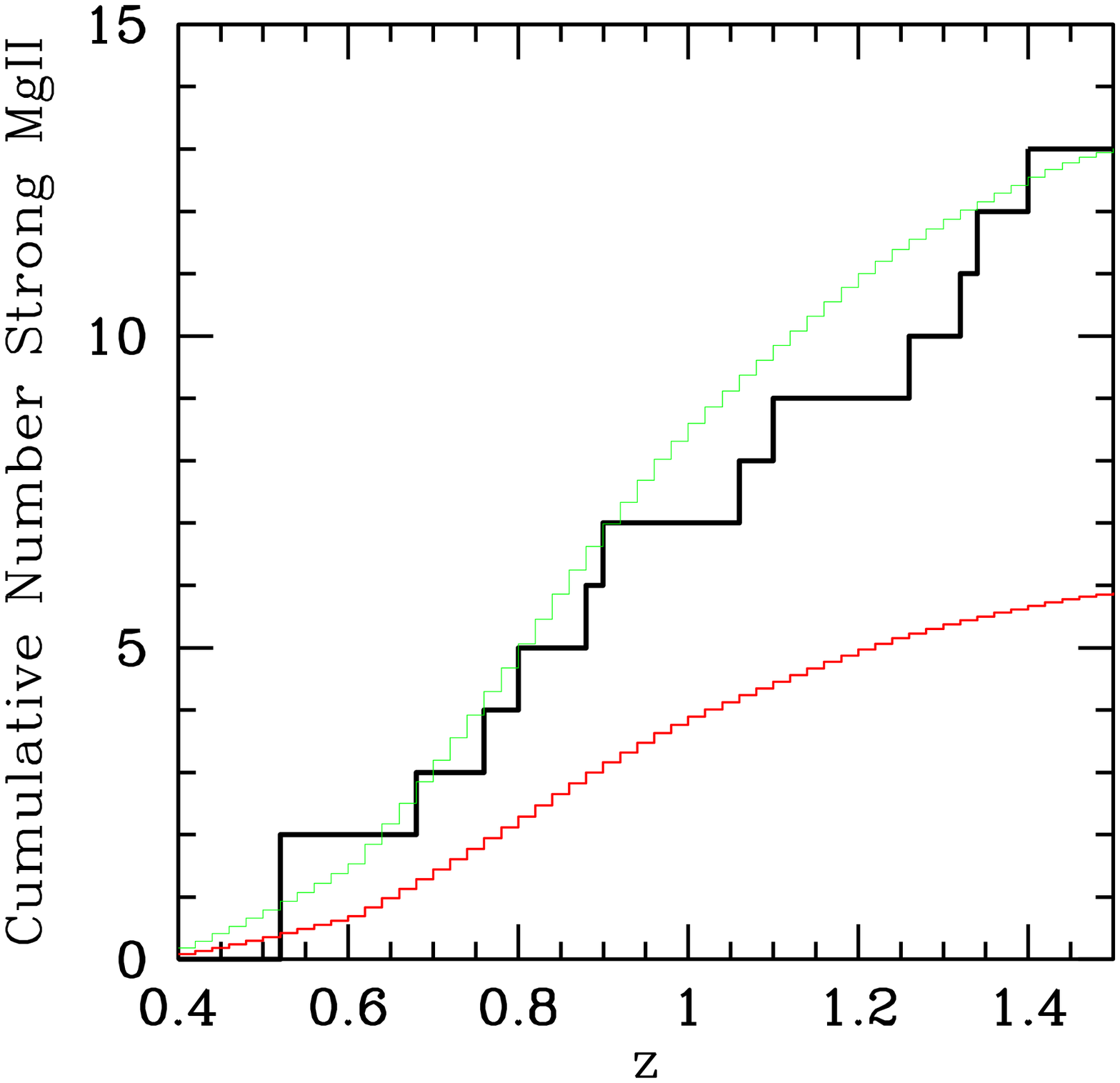}
      \includegraphics[width=6cm]{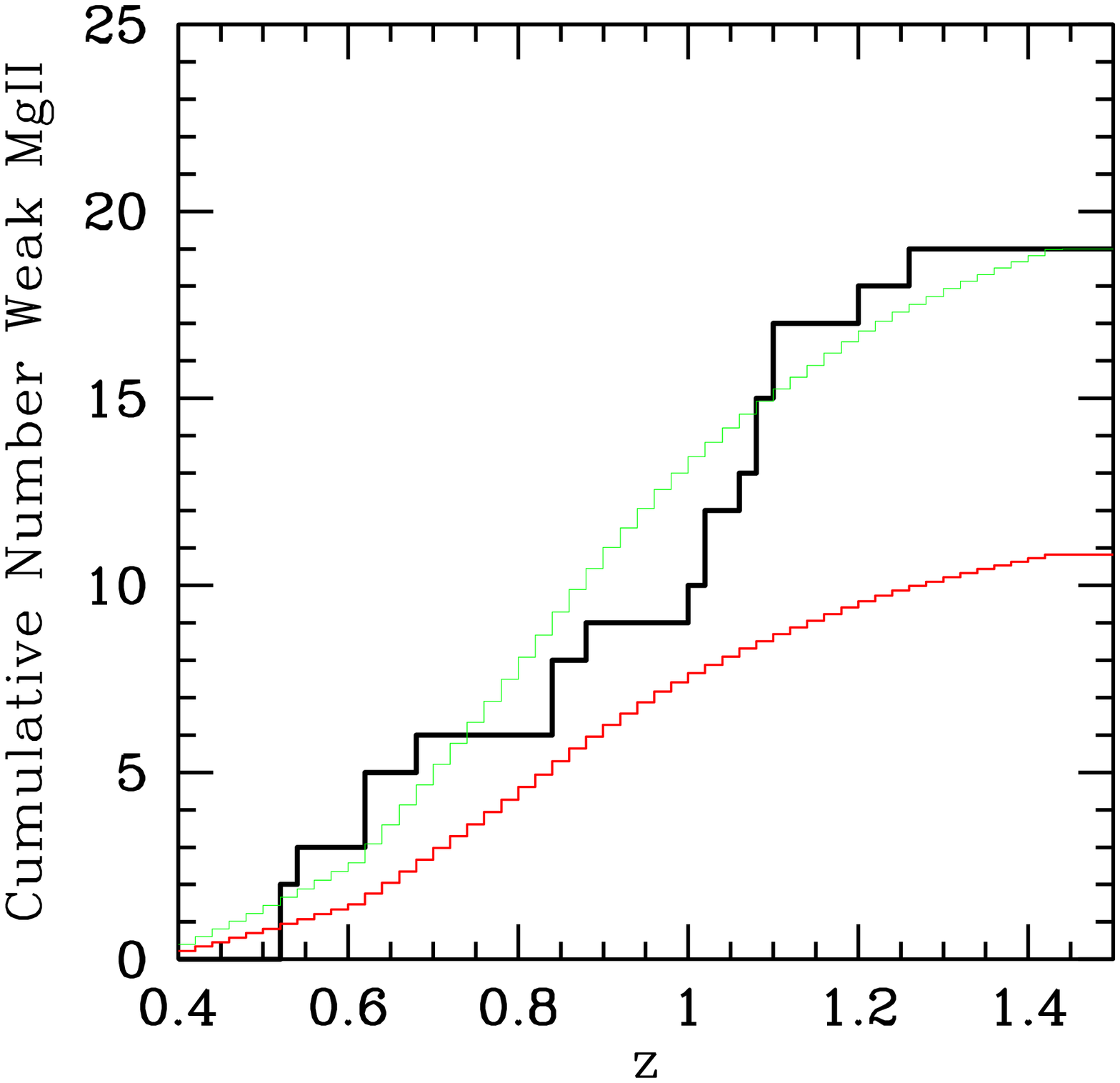}	
      \caption{\emph{Left:} Cumulative number of $w_{\rm r}(2796)$$>$$0.3$~\AA\ 	
               intervening Mg\,{\sc ii} systems towards Blazars, excluding the 	
              systems with $\Delta v < 5000$ km s$^{-1}$  (thick, black curve). 
               The curve of intermediate thickness (red) shows the cumulative  
             number  of such Mg\,{\sc ii} systems towards QSOs, adopting  the incidence 
             of these systems given by Nestor (see text), and the thinner (green)
              one is obtained by normalizing the latter to the total 
              number of Mg\,{\sc ii} systems towards Blazars. 
              {\emph{Center:} incidence of strong ($w_{\rm r}(2796)$$>$$1.0$~\AA)  Mg\,{\sc ii} systems
              adopting  the incidence of these systems given by Prochter et al. 
             (\cite{prochtera}). }
              {\emph{Right:} incidence of weak ($0.3< w_{\rm r}(2796) < 1.0$~\AA)  Mg\,{\sc ii} systems.              
               }}
         \label{cumz}
   \end{figure*}	

\subsection{ Global excess}

We measure the incidence of \MgII\ absorbers using our statistical 
sample of 45 Blazar sightlines.
The number density per unit redshift is $dN/dz=$ N$_{\rm obs}/\Delta z$, 
where N$_{\rm obs}$ is the number of observed absorption systems within 
the redshift path $\Delta z$. 

We first consider the entire \MgII\ absorber sample, i.e. all systems 
with $w_{\rm r}(2796)$$>$0.3~\AA. One of the strong system given in 
Table \ref{MgP8081} is excluded (northern Blazar) since the low S/N prevents  
detection of weak systems. 
We thus get: 
\begin{equation}
\frac{dN}{dz} (w_{\rm r}(2796)>0.3\,\AA)=\frac{31}{25.11}=1.23 ^{+0.26}_{-0.22} \mbox{~at~}  \langle z \rangle\simeq0.83\,. 
\end{equation}
The errors are based on Poisson statistics for small numbers with
limits corresponding to 1$\sigma$ confidence level of a Gaussian
distribution as tabulated by Gehrels (\cite{gehrels}) and also 
adopted by Tejos et al. (\cite{tejos09}).
To compare this value to the incidence of \MgII\ absorbers measured along 
QSO sight lines, we use the incidence of these systems per unit redshift 
derived from SDSS data. For $w_{\rm r}(2796) \geq 0.3$~\AA\, 
the incidence measured in QSO spectra is given by
Nestor from his analysis of SDSS DR4 (private communication): 
\begin{equation}
\frac {dN}{dz} = 0.288 + 0.4869 z - 0.0914 z^2.
\end{equation}

In Fig. \ref{cumz} (left panel), we present the cumulative number of \MgII\ Blazar 
systems versus redshift for the entire sample. 
 We also plot the cumulative distribution expected for \MgII\ QSO systems 
(computed using Eq. 2) 
and then adjust the normalization for comparison with Blazars. There is no evidence 
for a difference in functional form between Blazars and QSOs. 

Assuming that the Mg\,{\sc ii} Blazar absorbers are distributed at random 
along sightlines, as demonstrated for Mg\,{\sc ii} QSO absorbers with  
$\Delta v >  5000$ km  s$^{-1}$ (Steidel \&  Sargent \cite{steidel}; Wild et al. 
\cite{wild}),  the probability distribution for N, the number of systems 
expected in our  set of sightlines, should follow a Poisson distribution:
\begin{equation}
P({\rm N}_{\rm obs} \ge N_0, \mu) = \sum_{{\rm N}={\rm N_0}}^{\infty} \frac
{\mu^{\rm N} e^{-\mu}}{\rm N !},  
\label{eq:poisson}
\end{equation}
 where $\mu$ is the mean cumulative number expected for QSO systems over 
the whole set of Blazar sightlines (the value is given by the endpoint of the red curve
in Fig. \ref{cumz}). 
With $\mu = 15.98$ and $N_0=31$, we find $P= 0.00055$, indicating that similar 
$dN/dz$ for Blazars and QSOs are ruled out at a confidence level of 99.9\%.

The excess of  Mg\,{\sc ii} Blazar systems with $w_{\rm r}(2796) \geq 0.3$~\AA\ equals: 
\begin{equation}
Ex = \left( \frac {dN}{dz} \right)_{\rm Blazar}  /  \left( \frac {dN}{dz} \right)_{\rm QSO} = 2.0 \pm^{0.4}_{0.3},
\end{equation}
therefore providing a new piece of evidence that the observed incidence of \MgII\ absorbers depends on the type of background source. 
This value is similar to that reported for GRB lines-of-sight.

\subsection{Dependence on rest equivalent width}

In order to investigate the origin of the excess, we measure the \MgII\ 
incidence towards Blazars
for strong  ($w_{\rm r}(2796) > 1.0$~\AA) Mg\,{\sc ii} systems. We get: 
\begin{equation}
\frac{dN}{dz}=\frac{13}{29.93}=0.43 \pm^{0.16}_{0.12} \mbox{at} \langle z \rangle =0.82\,. 
\end{equation}

We adopt the fit  given by Prochter et al. (\cite{prochtera}) from their 
analysis of SDSS DR4: 
\begin{equation}
\frac{dN}{dz} = -0.026 + 0.374 z - 0.145 z^2 + 0.026  z^3. 
\end{equation}
While we detect N$_{\rm obs} =13$ systems, the number expected for QSO 
sightlines with the same $z$ coverage is only 5.88. Using Eq.~\ref{eq:poisson} with 
$\mu = 5.88$ and $N_0=13$, we get $P= 0.0075$, indicating that similar $dN/dz$ for 
Blazars and QSOs are ruled out at a confidence level of 99\%. 

The excess of  strong Mg\,{\sc ii} Blazar systems equals: 
\begin{equation}
Ex(w_{\rm r}(2796) > 1.0~\AA) = 2.2 \pm^{0.8}_{0.6}.
\end{equation}
This excess is similar to that found for GRBs (Tejos et al. \cite{tejos09}; Vergani et al. 
\cite{vergani}).
In Fig. \ref{cumz} (center panel), we show the cumulative redshift distribution for 
the strong Mg\,{\sc ii} Blazar systems. We also plot the distribution expected for 
Mg\,{\sc ii} QSO systems (Prochter et al. \cite{prochtera}) and that obtained after 
normalization to the total number of systems observed towards Blazars.  
%

For weak ($0.3 < w_{\rm r}(2796) < 1.0$~\AA) Mg\,{\sc ii} systems, 
we get:
\begin{equation}
\frac{dN}{dz}=\frac{19}{25.11}=0.76 \pm^{0.22}_{0.17} \mbox{at~} \langle z \rangle =0.83\,.
\end{equation}
Using Eq.~\ref{eq:poisson} with $N_0=19$ and the number expected for QSO sightlines with 
the same $\Delta z$, $\mu = 10.83$, we get $P= 0.015$. Thus similar $dN/dz$ for Blazars and 
QSOs are ruled out at a confidence level of 98.5\%, lower than for strong \MgII\ 
systems (the incidence of weak Mg\,{\sc ii} QSO systems is given by the difference in the values of 
$dN/dz$ at $\langle z \rangle =0.83$ obtained with Eqs. 2 and 6). 

The excess of  weak Mg\,{\sc ii} Blazar systems then equals: 
\begin{equation}
Ex(0.3 < w_{\rm r}(2796) < 1.0~\AA) = 1.7 \pm^{0.5}_{0.4}.
\end{equation}

\begin{figure}	
   \centering		
      \includegraphics[width=9cm]{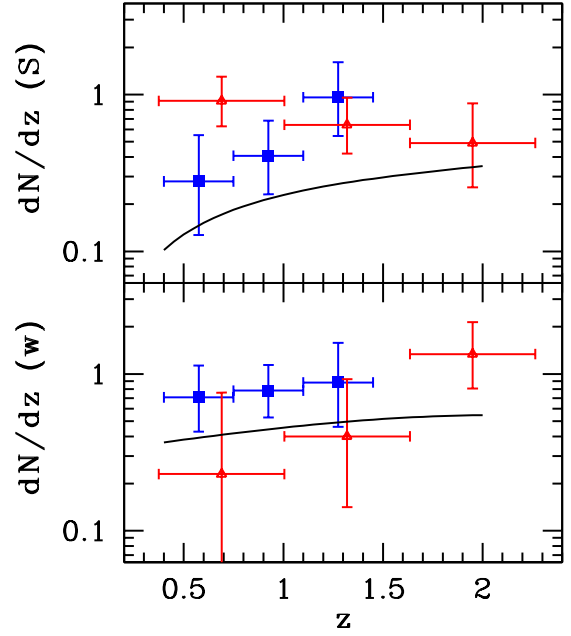}	
      \caption{
           Number density evolution of strong, $w_{\rm r}(2796) > 1.0$~\AA\ 	
           (top panel), and weak, $0.3<w_{\rm r}(2796) < 1.0$~\AA\  (bottom panel), 
          Mg\,{\sc ii} systems towards Blazars (blue squares), GRBs 
           (red small triangles) as reported by Vergani et al. (\cite{vergani}) 
             for their overall sample (with a downwards correction for the absolute 
           values of strong systems: Vergani, private communication)   
            and SDSS QSOs (solid line) as given by Prochter et al. (2006) for strong 
            systems and Nestor (private communication) for weaker systems.} 
         \label{dNdz}	
   \end{figure}

Finally we note that our analysis \emph{indicates} a possible difference in the 
redshift evolution of the excess of Blazar and GRB strong absorbers. 
The excess of strong systems is higher at high $z_{\rm abs}$ along Blazar
sightlines (a linear fit yields $dN/dz \approx 1.0 z - 0.4$) 
whereas the opposite trend is seen towards GRBs (Vergani et al. 
\cite{vergani}).  No difference is seen for weak systems. 
The evolutions in redshift are  shown in Fig.~\ref{dNdz} for 
strong and weak Mg\,{\sc ii} systems.

%



\subsection{ Results for BL Lacs only}			

 Since 35 out of the 45 targets in the overall Blazar sample belong to the BL Lac 
class, we can determine the incidence of Mg\,{\sc ii} BL Lac systems only, with still 
statistically meaningfull  results. 

As in Sect. 4.1, we first consider the entire \MgII\ absorber sample (24 systems 
with $w_{\rm r}(2796)>0.3$~\AA). This yields:  
\begin{equation}
\frac{dN}{dz} (w_{\rm r}(2796)>0.3\,\AA)=\frac{24}{18.33}=1.31 ^{+0.33}_{-0.26} \mbox{~at~}  \langle z \rangle\simeq0.84\,. 
\end{equation} 
For the 10 strong ($w_{\rm r}(2796)>1.0$~\AA) and 15 weak ($0.3<w_{\rm r}(2796)<1.0$~\AA)
Mg\,{\sc ii} BL Lac systems separately, we get: 
\begin{equation}
\frac{dN}{dz}(strong)=\frac{10}{22.81}=0.44 \pm^{0.19}_{0.14} \mbox{at} \langle z \rangle =0.82\,. 
\end{equation}
\begin{equation}
\frac{dN}{dz} (weak)=\frac{15}{18.33}=0.82 \pm^{0.27}_{0.21} \mbox{at~} \langle z \rangle =0.84\,.
\end{equation}
These results are nearly identical to those obtained for the whole Blazar samples (Eqs. 
1, 5 and 8), although with larger uncertainties. 
We thus find no evidence for a distinct behavior of BL Lac and FSRQ sightlines, which a 
posteriori justifies considering the Blazar sample as a whole. 

The excess of Mg\,{\sc ii} BL Lac systems equals: 
\begin{equation}
Ex = 2.1 \pm^{0.5}_{0.4},  2.2 \pm^{0.9}_{0.7}, 1.9 \pm^{0.6}_{0.5}, 
\end{equation}
for the entire, strong and weak samples, respectively, while Stocke \& Rector (\cite{stocke}) 
found on the basis of ten sightlines an excess of about 4 - 5 (note that among their five 
strong systems one is associated and another one is not confirmed).	

\subsection{Relative velocity distribution}	
 	
We now test whether the excess of strong  Mg\,{\sc ii} Blazar systems arises from 
preferential values of relative velocity between the Blazar and the absorber. We 
 consider the cumulative distribution of Mg\,{\sc ii} systems as a function of 
$\beta$, 
\begin{equation}
\beta \equiv \frac{v}{\rm c}  = \frac {(1+z_{\rm em})^2-(1+z_{\rm abs})^2}
{(1+z_{\rm em})^2+(1+z_{\rm abs})^2},  
\end{equation}
where $v$  is due to the expansion of the Universe and the peculiar velocity of gas ejected 
by a source at $z_{\rm em}$. 

\begin{figure*}	
   \centering		
      \includegraphics[width=6cm]{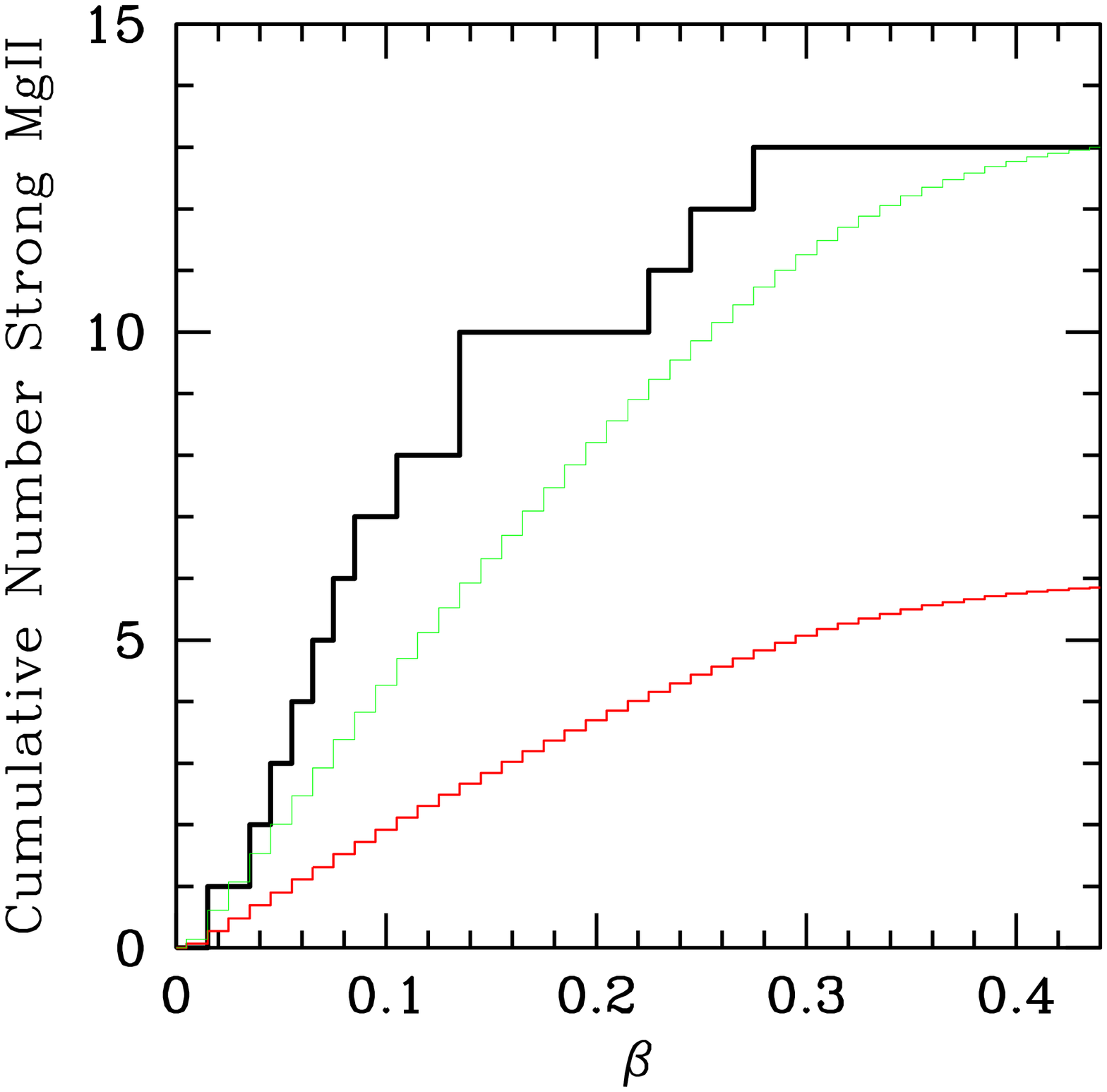}
      \hspace{2cm}    
     \includegraphics[width=6cm]{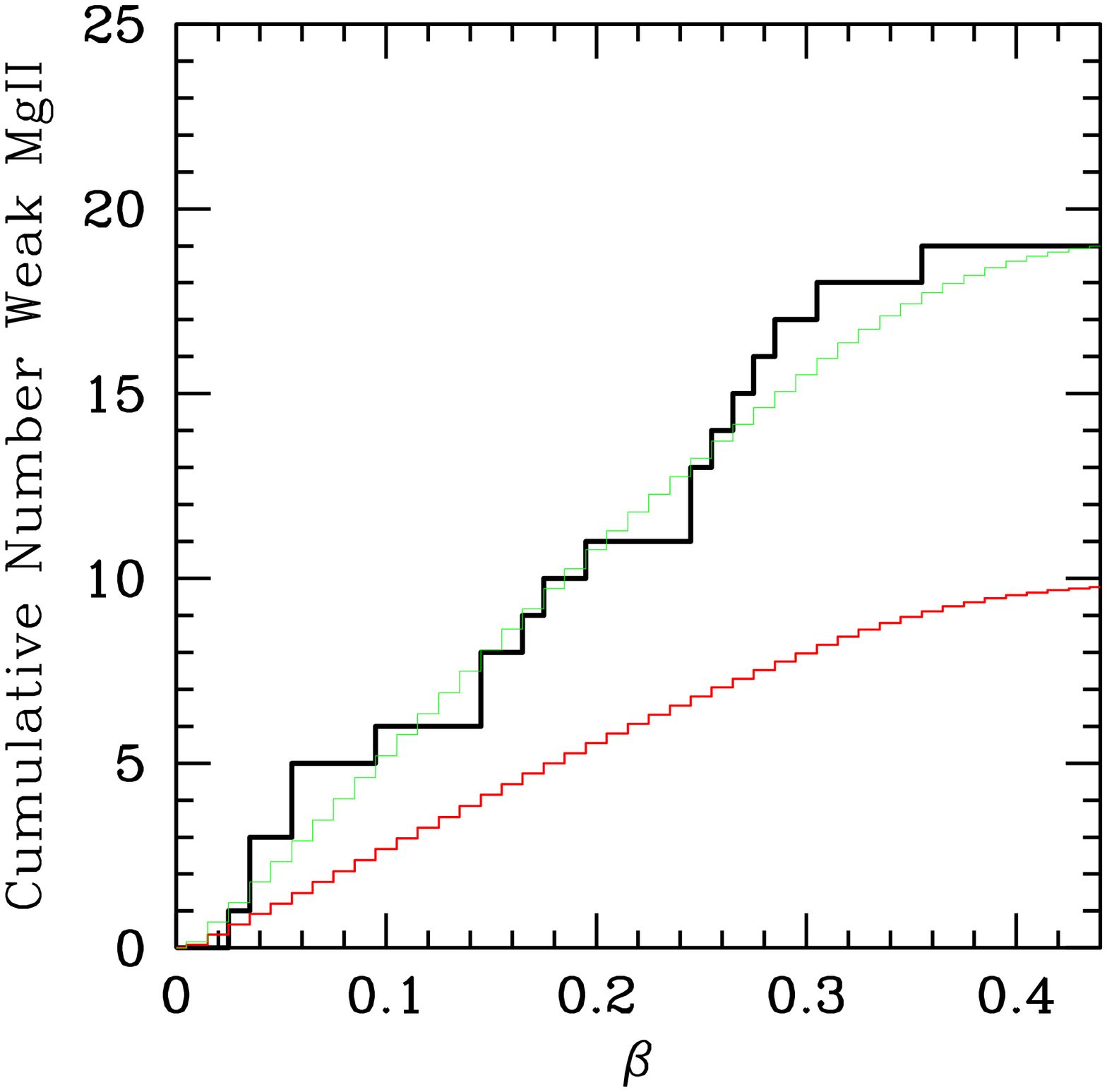}	
      \caption{\emph{Left:} Cumulative number of strong ($w_{\rm r}(2796) > 1.0$~\AA) 	
               intervening Mg\,{\sc ii} systems towards Blazars versus 
              $\beta \equiv v/$c, excluding the systems with 	
              $\Delta v (z_{\rm em}, z_{\rm abs}) < 5000$ km s$^{-1}$ or 
               $\beta < 0.0167$	(thicker, black curve).
               The curve of intermediate thickness (red) shows the predicted number 
              of strong Mg\,{\sc ii} systems, adopting	the incidence 	
             of these systems towards QSOs derived by Prochter et al. 
             (\cite{prochtera}), and the thinner (green) 
              one is obtained by normalizing the latter to the total 
              number of strong Mg\,{\sc ii} systems towards Blazars. 
	\emph{Right:}  Same for weak ($0.3<w_{\rm r}(2796)< 1.0$~\AA) 	
                Mg\,{\sc ii} systems. 
}
         \label{beta}	
	
   \end{figure*}

In Fig. \ref{beta}, we give the cumulative
$\beta$ distribution for strong and weak systems respectively. For comparison, 
we need to compute the distribution expected in $\beta$-space from the background 
population of intervening Mg\,{\sc ii} absorbers. To this aim, we consider the redshift 
distribution adopted in Sect. 4.1 and build the $\beta$ distribution for each 
 sightline (due to varying $z_{\rm em}$ values from one target to another, the 
transformation $z \rightarrow \beta$ is specific to each sightline). Next, for 
any $\beta$ bin, we combine all sightlines contributing to this bin.  
As expected, the same excess of systems as in Fig. \ref{cumz}  is apparent. 
More significantly, after scaling to the same total number of systems 
to assess whether the {\it shapes} of the distributions are similar, one can see that 
strong systems display a relative excess of low $\beta \sim 0.06$ to 0.18 systems while 
the shapes of the QSO and Blazar distribution for weak systems are in quite 
good agreement. To estimate the significance of the excess of strong systems at low 
$\beta$, we perform a KS test and find that we can reject the null hypothesis of a 
similar underlying distribution only at the 80\% 
significance level.  This result must then be considered only as a trend. 

\subsection{Associated systems}	

Our results allow us to compare the number of  Mg\,{\sc ii} systems	
at $z_{\rm abs} \approx z_{\rm em}$ to that found for QSOs; the	
latter is now well constrained by studies based on SDDS data 
(Wild et al. \cite{wild}). 	
Associated systems in Blazars might be more numerous than in QSOs 	
due e.g. to a richer environment  or to differing jet properties. 
	
At least two associated systems with $w_{\rm r}(2796) > 0.3$~\AA\ are present; 
this is a lower limit because of the sightlines for which no emission redshift 
is available. The number density of $w_{\rm r}$$>$$0.3$~\AA\ QSO systems 
at $z=1$ is $dN/dz=0.68$ which implies an expected number of associated systems 
per sightline of 
\begin{equation}
\Delta N =  \frac{dN}{dz} \,\,\, |\frac {dz} {d\beta}| 
\,\,\, \Delta\beta = 0.023, 
\end{equation}
using 
\begin{equation}
|\frac {dz} {d\beta}| \simeq (1+z) = 2. 
\end{equation}
Thus, for the 41 sightlines in which we can detect $\beta \simeq 0$ systems, 
we expect $\sim$0.9 system. Towards radio-quiet QSOs, the fraction of 
$w_{\rm r}(2796)>0.5$~\AA\ systems at $\Delta v<5000$ km s$^{-1}$, 
relative to the background level,  is 1.7 while this factor reaches 3.3 
towards radio-loud QSOs (Wild et al. \cite{wild}; Wild, private communication: 
updated results based on the SDSS DR5 that are fairly insensitive to the 
$w_{\rm r}$ limit). Thus, if Blazars have a similar excess of 
associated systems than  radio-loud QSOs, we expect 3.3 associated systems at 
$\Delta v<5000$ km s$^{-1}$ in our sample whereas there are two associated and 
three possibly associated systems. The latter could then be truly associated  
systems that indeed define the  BL Lac/Blazar emission redshifts.

\section{Discussion}	
%

\subsection {Comparison between Blazars, GRBs and QSOs}	
%

Our analysis provides a new piece of evidence that the observed incidence of \MgII\ 
absorbers  depends on the type of background source. Towards both Blazars 
and GRBs, $z\sim1$ 
\MgII\ absorbers are found in excess by roughly a factor of 2 with respect to QSO 
sightlines. In each case, the difference is significant at the $\sim3\sigma$ level. 


	
The reported excesses for GRB and Blazar sightlines are similar for \MgII\ absorbers with $W_0>1\,{\rm\AA}$.
For weak systems, comparison of the results obtained for Blazars and GRBs 
is less straightforward due to small number statistics for GRBs. Whereas 
we find an excess towards Blazars equal to 1.8$\pm^{0.5}_{0.4}$  at 
$\langle z \rangle =0.83$  (2.5$\sigma$ significance level), there is 
no significant excess for GRB absorbers (Vergani et al. \cite{vergani}) 
with a  value of 1.2$\pm^{0.6}_{0.4}$  (using Poisson statistics for the 
errors) at $\langle z \rangle = 1.57$. Nevertheless, the GRB results differ 
from the Blazar ones at only a 1$\sigma$ significance level and it can  
 be seen in Fig. \ref{dNdz} that the GRB absorber excess at their highest  
$z$ bin follows the trend found at lower $z$ for weak Mg\,{\sc ii}  Blazar 
absorbers. It is thus not really possible to conclude whether or not there is a 
difference between the weak Mg\,{\sc ii} Blazar and GBR absorber populations.  

Our analysis also suggests a possible difference in the redshift evolution 
of this excess but larger samples are needed to make definite statements.

\subsection{Dust and amplification biases}

Dust extinction and/or gravitational magnification due to foreground
matter can modify the observed density of background sources and therefore 
the observed incidence of intervening absorbers. Several authors already 
pointed out these effects as potentially responsible for (some of) the 
\MgII\ absorber excess in GRB spectra (Porciani et al. \cite{porciani}). 
Here we explain why these effects are unlikely to have a significant
contribution in giving rise to an apparent excess of strong \MgII\ absorbers 
both in GRB and Blazars spectra compared to the value measured with QSOs.

GRBs are preselected on the basis of their gamma-ray flux. Rapid follow-up 
observations of the afterglow emission allow us to observe these objects while 
their apparent magnitude is significantly (i.e. several magnitudes) brighter 
than typical limiting magnitudes. This difference is much larger than the 
typical levels of extinction induced by intervening \MgII\ absorbers.
The dust content of these systems has been quantified by M\'enard et
al. (\cite{menard}): these authors showed that on average, \MgII\
absorbers with $w_{\rm r}(2796)\sim1\RAA$
give rise to a color excess $E(B-V)\simeq0.01$ mag.
These authors estimated that about one percent of SDSS QSOs are missed
due to the presence of dust in foreground \MgII\ absorbers.
Most BL Lac objects are identified on the basis of their radio or
X-ray emission (e.g. Padovani \& Giommi \cite{padovani95})
or their gamma-ray emission. This preselection is not sensitive to a
dust obscuration bias (only low frequency X-rays would be affected by
absorption but very large hydrogen column densities are required). The
optical follow-up could be affected by such an effect. However,  Blazars are 
variable in this wavelength range with a typical amplitude reaching about one 
magnitude,  and   for a few Blazars of our sample up to 4-5 magnitudes. 
 Consequently, as in GRBs detection of \MgII\ absorbers towards Blazars 
is not affected by dust extinction. 

Gravitational magnification can change the density of background
sources. This effect depends on the gradient of the matter over-density 
along the sightlines and the slope of the source number counts as a 
function of magnitude (Narayan \cite{nara}):
\begin{equation}
n(m) = \mu^{\alpha(m)-1}\,n_o(m)\,,
\end{equation}
where $n$ and $n_0$ are the lensed and unlensed density of sources in
the sky, $\mu$ is the gravitational magnification along their
sightline and $\alpha(m)$ is the slope of the source number counts as
a function of magnitude. The average magnification effect induced by
galaxies was detected by Scranton et al. (\cite{scran}). These authors
report an average magnification excess of order one percent for impact
parameters of about 50 kpc. Gravitational lensing by absorbers  has
been theoretically addressed by a number of authors (Bartelmann \&
Loeb \cite{bartelmann}; M\'enard \cite{menard05}). Given the typical
impact parameters of strong \MgII\ absorbers ($\sim20$-50 kpc) the
magnification excess $\mu-1$ is expected to be small, i.e. of order
a few percents. Observationally, constraints on QSO magnification by
\MgII\ absorbers have been reported by M\'enard et al.
(\cite{menard}). These authors report $\mu-1<0.1$.
 This method cannot be applied to the Blazar population since their 
number counts as a function of magnitude is ill defined due to their strong 
optical variability. But, the latter being far larger than any expected 
gravitational magnification, any significant amplification bias is most 
unlikely for Blazars. Finally, HST imaging surveys show that 
gravitational lensing appears to be unimportant to the BL Lac phenomenon 
(Giovannini et al. \cite{giovannini}).
Recently Wyithe et al.  (\cite{wyithe}) proposed strong lensing as an
explanation for the higher incidence of \MgII\ absorbers in front of
GRBs. They argue that 20 to 60\% of the GRBs with strong \MgII\
absorbers are strongly lensed. Here we note that if this was the case,
the GRB host galaxy should also be strongly lensed and appear as
arcs or multiple images. So far HST imaging of GRB host galaxies have
not reported such a phenomenon (Conselice et al. \cite{conselice}), indicating 
that strong lensing is not often induced by intervening \MgII\ absorbers.

 Therefore, while dust extinction can lower the apparent incidence of
absorbers in front of quasars and gravitational lensing can increase it in
front of GRBs and Blazars, the amplitude of these effects is at least an
order of magnitude lower than what would be required to explain a factor
$\sim$2 change in the incidence of \MgII\ absorbers.

\subsection{An intrinsic origin?}

We now consider whether a significant fraction of the \MgII\ system excess 
could be intrinsic by exploring some of the physical conditions required 
for cold gas to be entrained at high velocities.
Such a scenario is appealing in the context of Blazars since these sources 
are known to display relativistic jets pointing towards us. If material is
present in the neighborhood (either gas surrounding the AGN, within the 
interstellar medium  of the host galaxy, or further away within its halo) it 
might be swept up by the jet and accelerated to high velocities. In our 
sample, there is no marked accumulation of systems at $\beta < 0.02$ thus, if 
such systems give a significant contribution ($\simeq$ 50\%) to the 
observed Mg\,{\sc ii} systems, ejection velocities of a few $\times\,10^{4}$ 
km s$^{-1}$ must be involved.

Towards QSOs, outflow velocities in this range are currently observed
for BAL systems. The latter are thought to arise when the line of
sight intersects the wind emanating from the accretion disk with the
appropriate inclination angle  - typically 30 deg (see Elvis \cite{elvis})
 - which accounts for the large velocity dispersion found in these
systems. Regarding Blazars, the inclination is supposed to be smaller
(around 10 deg;  Fan et al. \cite{fan09}) and as a result, the
line of sight is not expected to cross the disk wind. Indeed, BAL
absorption is usually not seen in Blazar spectra.  

One may rather invoke material impacted by the jet along its way as the
source of the absorption. This could be for instance diffuse 
gas located relatively close to the AGN. 
In such a configuration, jet material is expected to sweep up the 
external gas and while the corresponding column density ($N_e$) increases, 
the advance velocity of the interface ($V$) decreases (Falle \cite{falle}; 
Kaiser \& Alexander \cite{kaiser}; Komissarov \& Falle \cite{komi}). 
If we identify the advance velocity with the apparent
ejection velocity of the absorbing gas, the question is then: can we
still have total column densities of swept up gas as large as 
$10^{18-20}$cm$^{-2}$ together with velocities as high as 
 a few $10^4$ km s$^{-1}$? Komissarov \& Falle (\cite{komi}) have shown
 that in the early phase of a relativistic jet growth $V \simeq c$ while
 later, the dynamics of the interface follows a power-law behavior with
 respect to the time $t$, with the jet length ($L$) increasing as
 $t^{3/5}$, corresponding 
to a velocity $V=dL/dt$ decreasing as $t^{-2/5}$. In this regime, one
can use the power-law expressions provided by Komissarov \& Falle 
(\cite{komi}) to relate directly $L$ to $V$ (by eliminating $t$), which
leads to 
\begin{equation}
L =  \left(\frac{3}{5}\right)^{3/2} \left(\frac{P}{\rho_e}\right)^{1/2}  V^{-3/2}.
\end{equation}
\noindent
The column density of swept up gas, $N_e$, simply writes 
$N_e = L \times n_e$. Assuming for simplicity a pure hydrogen gas to
connect $\rho_e$ and $n_e$, we finally get the numerical relation: 
\begin{equation}
N_e(cm^{-2}) \simeq 2\times 10^{19}  \left( \frac {P}{10^{45} erg s^{-1}}\right)^{0.5}
\left( \frac{n_e}{0.1 cm^{-3}}\right)^{0.5} \left(\frac{V}{0.1c}\right)^{-1.5}.
\end{equation}
\noindent
This relation indicates that for reasonable jet powers and ambient gas
densities, a column density as large as $10^{18-20}$cm$^{-2}$ can be 
swept up, with a relative velocity attaining $0.1c$.
The above estimate for $N_e$ is an upper limit for the amount of gas
effectively present at the interface because part of the swept up
external gas flows away from the contact zone. However, even if only
a small fraction (e.g. 0.1) remains at the interface, the column density can 
be large enough to induce the observed absorption systems. 
 Note that such an intrinsic scenario does not necessarily imply a large velocity 
dispersion since, contrary to the case of QSO BAL systems, the absorbing material is 
not closely associated with the active nucleus.

\section{Conclusions}	
%
	


Attempting to shed light on the unexplained excess of \MgII\ absorbers in front of 
GRBs, we measured the incidence of such absorbers in front of another class of 
objects: the Blazars.

We observed with FORS1 at the ESO-VLT 42 Blazars with an emission redshift 
$0.8<z_{\rm em}<1.9$, to which we added the three high $z$ northern objects belonging 
to the 1Jy BL Lac sample. From this sample we detect 32 Mg\,{\sc ii} absorbers in 
the redshift range 0.35-1.45. Our main results are as follows.
\begin{enumerate}
\item We find an excess in the incidence of Mg\,{\sc ii} absorbers compared to that 
measured towards QSOs by a factor $\sim 2$, detected at 3$\sigma$. The amplitude of 
the effect is similar to that found along GRB sightlines. 
\item In front of Blazars, the excess is apparent for both "strong", i.e. 
$w_{\rm r}(2796)>1.0$~\AA, and weaker systems, with $0.3<w_{\rm r}(2796)<1.0$~\AA.
\item The dependence on velocity separation with respect to the background Blazars 
indicates, at the $\sim1.5\sigma$ level, a potential excess for $0.06<\beta<0.18$.
\end{enumerate}
This work brings another piece of evidence that the incidence of extragalactic absorbers 
systems might depend on the type of background source considered. We argue that biases 
involving dust extinction, weak and/or strong gravitational amplification are not likely 
to affect the frequency of Mg\,{\sc ii} systems detected towards either Blazars and GRBs 
at a significant level. We also briefly discuss the physical conditions required for 
these absorber systems to be physically associated with the Blazars.

We claim that a satisfactory explanation for the apparent excess of \MgII\ 
absorbers in front of certain types of background sources still awaits. 
Further investigations should be made on both theoretical and observational 
sides. For the former, realistic numerical modelling of the jet/ambient gas 
interaction should focus not only on jet properties as commonly done (see e.g. 
Choi et al. \cite{choi})  but rather on those of the swept up gas at different 
stages of the jet growth. We stress in particular that taking into account the 
thermal behavior of the gas (heating/cooling processes) is important to properly 
assess its physical properties (ionisation level, spatial structure, 
temperature; e.g.  Mellema et al. \cite{mell}). 

Given the lack of current theoretical understanding, more observational results 
might offer us some guidance, especially by exploring dependencies with source 
properties. 
A first test is to split QSOs in various sub-classes. The most straightforward  
one involves their radio brightness: this has been investigated by Ellison et al. 
(\cite{ellison}) who have shown that the incidence of \MgII\ absorbers does not appear 
to depend on the radio loudness of the background QSO. 
Other tests concern the size of the optical emission regions. For QSOs, 
Pontzen et al. (\cite{pontzen}) reported a consistent \MgII\ incidence measured 
over the continuum regions and the larger, C\,{\sc iii} emission line regions. 
Such a test should also be conducted for Blazars in high and low states, 
when the continuum flux originates mainly either from the jet or from the accretion 
disk, respectively. Those Blazars from our sample displaying several systems are 
obvious targets for such a study. Repeat high resolution spectra would in 
particular  constrain possible variability of various absorption lines 
(Mg\,{\sc ii}, Mg\,{\sc i}, Fe\,{\sc ii}), thus test the 
intrinsic origin scenario, and yield kinematics and ionization properties of 
\MgII\ Blazar absorbers to be compared with those of \MgII\ QSO absorbers.






\begin{acknowledgements}	
 We thank R. M. Plotkin for providing information prior to publication and 
the anonymous referee for his/her useful comments and suggestions. 
We are also grateful to F. Daigne and S. Falle for helpful discussions on 
jet models. 
\end{acknowledgements}

\Online

\onecolumn

\begin{appendix}

\section{New emission redshifts}	
%

\begin{figure*}[ht]	
   \centering	
      \includegraphics[height=21cm,angle=-90.]{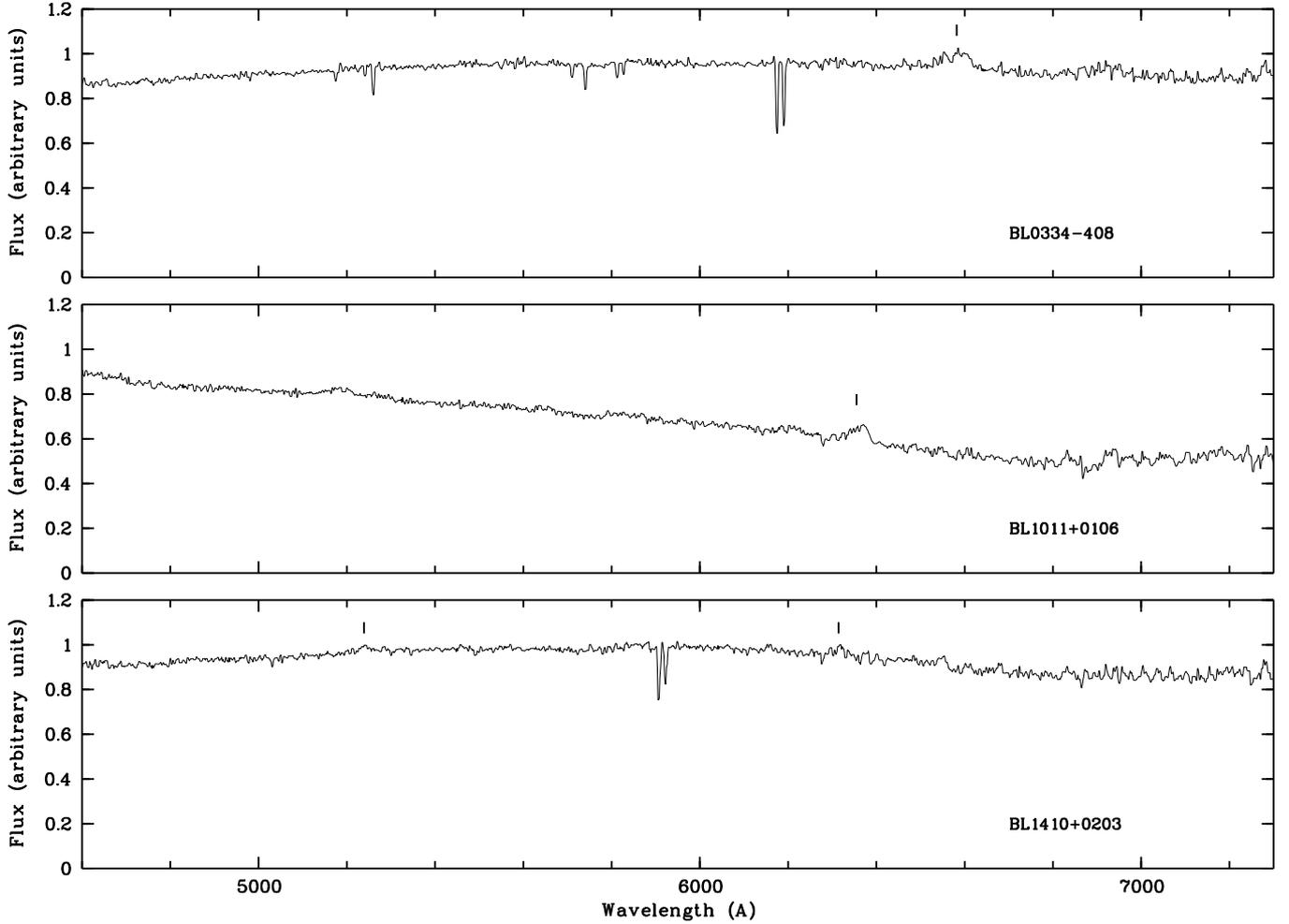}	
    \caption{Blazars with new emission redshift
   (Mg\,{\sc ii} and C\,{\sc ii}] emission lines).}
         \label{newzem}	
   \end{figure*}

The spectra of three Blazars show very weak emission lines yielding new emission 
redshifts. They are presented in Fig. \ref{newzem}. 

	
In PKS 0332$-$403 (Fig. \ref{newzem} top panel), the weak emission at 
$\lambda = 6580$ \AA\ is identified as Mg\,{\sc ii} at $z_{\rm em}=1.351$. As 
reported in Table \ref{MgP8081}, there is a weak Mg\,{\sc ii} absorption doublet at 
$z_{\rm abs}=1.2083$ with associated Fe\,{\sc ii} absorptions (the  $\lambda2600$  
doublet and the $\lambda2382$ triplet); the other Mg\,{\sc ii} system at 
$z_{\rm abs}=1.0791$ is too weak to be included in our statistical Mg\,{\sc ii} sample. 

In PKS 1008$+$013 (Fig. \ref{newzem} central panel),  there is an asymmetric emission 
line centred at $\lambda = 6368$ \AA\ identified as Mg\,{\sc ii} at $z_{\rm em}=1.275$, which 
is also clearly present in the SDSS spectrum of this source. We thus do not confirm 
the 'uncertain' redshift, $z_{\rm em}=0.8615$, given by Collinge et al. (\cite{collinge}). 
We do not detect any absorption line in the FORS1 spectrum of this Blazar. 

In PKS 1407$+$022 (Fig. \ref{newzem} bottom panel), there are two weak emission lines 
at $\lambda = 5238$ and 6308 \AA\ that we identified as  C\,{\sc ii}] 
and Mg\,{\sc ii} at $z_{\rm em}=1.253$. There is a weak  Mg\,{\sc ii} absorption doublet
at $z_{\rm abs}=1.1123$ (see Table \ref{MgP8081}) with possible associated Fe\,{\sc ii} 
absorptions (marginal detections of Fe\,{\sc ii}2382 and Fe\,{\sc ii}2600). 

\section{The discovery of a broad absorption line in a Blazar spectrum}

	
   \begin{figure*}	
   \centering	
      \includegraphics[height=26.cm,angle=-90.]{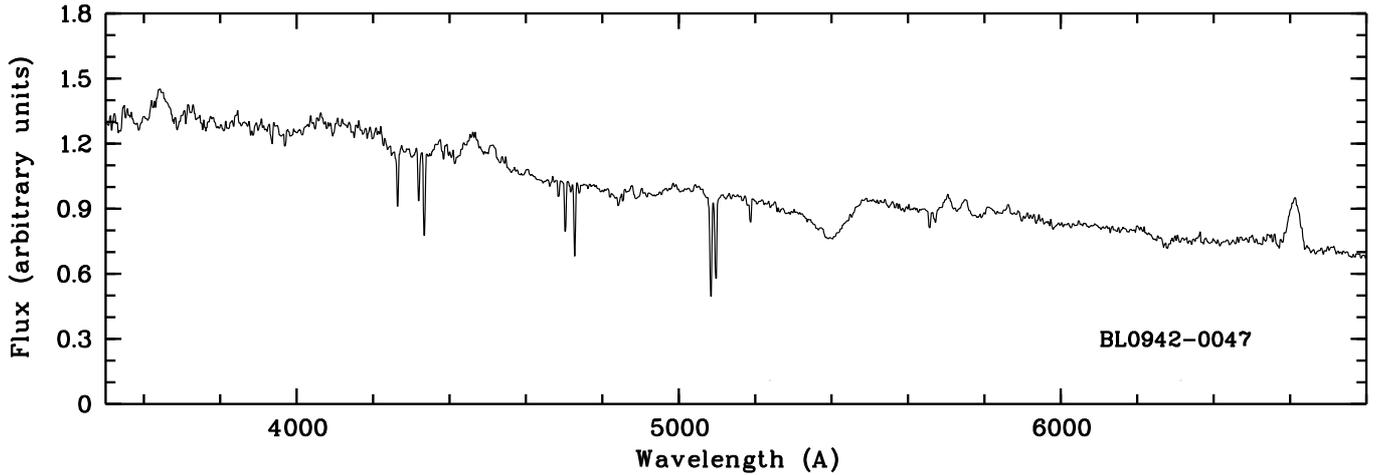}		
    \caption{BL0942-0047 with C\,{\sc iv} and Mg\,{\sc ii}  emission lines
   and a Mg\,{\sc ii} BAL at $\lambda \sim 5400$ \AA.}
         \label{0942}	
   \end{figure*}	

For one Blazar (the weak FIRST radio source SDSS J094257.8$-$004705), there is 
an unusual broad absorption line with a FWHM $\simeq 6000$ km s$^{-1}$ 
as shown in Fig. \ref{0942}. This feature cannot be the signature of a dust 
shallow absorption at 2175 \AA\ associated with the Blazar ($z_{\rm em}=1.362$), 
since it should be at $\lambda_{\rm r}>2285$\AA. Thus, it is identified as  
a Mg\,{\sc ii} BAL at $z_{\rm abs} = 0.929$ or $\Delta v= 0.20$c. 
Such highly detached BALs, although very rare, have already been detected at low 
redshift (e.g. the high ionization BAL in PG 2302+029 at $z_{\rm abs} = 0.695$ 
reported by Jannuzi et al. (\cite{jannuzi})). 
In sources at 
$1.5 \leq z \leq 3$ from the FIRST Bright Quasar Survey (FBQS), there is a 
5\% fraction of low ionization (Lo) BALs (Becker et al. \cite{becker}); the 
frequency of LoBAL QSOs is dependent of radio-loudness and decreases for the 
most radio-luminous QSOs. We thus expect at most two LoBAL QSOs in our Blazar 
sample since 33 of the 45 Blazars are strong ($\sim$1 Jy) radio sources, 
which is consistent with our single LoBAL detection. 


There are also two narrow absorption systems in the Blazar spectrum. The strong 
Mg\,{\sc ii} doublet at $z_{\rm abs}= 0.8182$ has Mg\,{\sc i}, Mn\,{\sc ii} 
and Fe\,{\sc ii} associated strong absorptions. 
The weak system at $z_{\rm abs}= 1.0231$ shows Mg\,{\sc ii} absorption only. 

\section{Not confirmed strong  Mg\,{\sc ii} absorption in one source of 
the Stocke \& Rector BL Lac sample}	
%

PKS 0426$-$380 is one of the ten sources presented by Stocke \& Rector (\cite{stocke}),  
sample for which an excess of strong  Mg\,{\sc ii} in BL~Lac objects had first been 
reported. The high S/N (270) of our  FORS1 spectrum enables us to detect very weak 
absorption lines. The absorber at $z_{\rm abs}= 1.0283$ is not confirmed as a 
strong Mg\,{\sc ii} system (w$_{\rm r}(2796) = 0.56$~\AA) and has weak Fe\,{\sc ii}(2382,2600) 
and Al\,{\sc iii} doublet associated absorptions. 

There is a stronger Mg\,{\sc ii} doublet 
(w$_{\rm r}(2796) = 0.93$~\AA)  at $z_{\rm abs}= 0.5592$ with  associated Mg\,{\sc i} 
absorption together with  the five stronger Fe\,{\sc ii} lines. 
We also detect Galactic Ca\,{\sc ii} absorption, as well as the only intervening 
($z_{\rm abs} > 0$) Ca\,{\sc ii} absorber in our Blazar sample. The latter is at 
$z_{\rm abs} = 0.1940$ with a marginal detection of the weaker line of the doublet 
but exactly at the same redshift. 

	
   \begin{figure*}[hb]	
   \centering	
      \includegraphics[height=26.cm,angle=-90.]{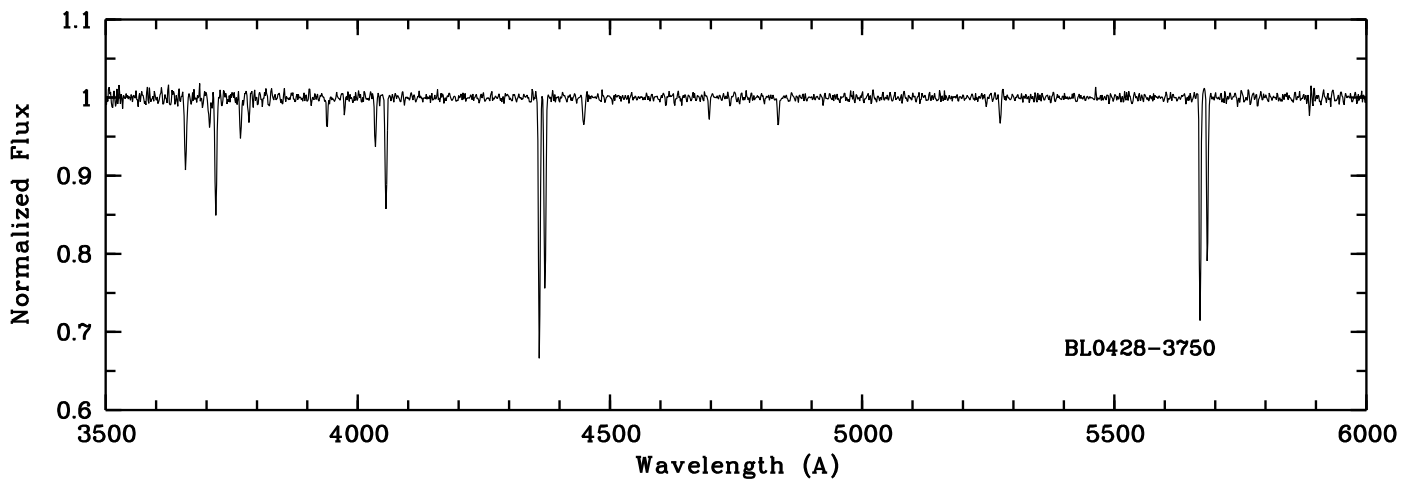}
    \caption{In this BL Lac the Mg\,{\sc ii} absorption at $z_{\rm abs}= 1.0283$ is 
    weaker than reported by Stocke \& Rector (\cite{stocke}).
}
         \label{BL0428}	
   \end{figure*}	

\section {Known and potential DLAs}	
%
Classical Damped Ly$\alpha$ systems (DLAs) have  H\,{\sc i} column densities 
N(H\,{\sc i}) $> 2\times 10^{20}$ cm$^{-2}$. The fraction of DLAs among 
intervening QSO absorbers, with constraints on Mg\,{\sc ii}2796 and 
Fe\,{\sc ii}2600 (or the ratio of their equivalent widths) and on Mg\,{\sc i}2852 
is about 40\% (Rao et al. \cite{rao}). Is the DLA fraction similar for Blazars? 
Among our sample of 45 Blazars, N(H\,{\sc i}) can be derived from Ly$\alpha$ 
absorption line profiles for eight sources only, i.e. those with HST/FOS or STIS 
spectra. Since most of our selected Blazars are strong radio sources, the DLA 
subpopulation which traces a cold gas phase can also be identified by 21~cm 
absorption. 

Three of the strong Mg\,{\sc ii} Blazar absorbers have UV spectra. One is a DLA, 
the  $z_{\rm abs}=0.5245$ absorber towards PKS 0235$+$164, 
and two are sub-DLAs with N(H\,{\sc i})  $< 1 \times 10^{20}$ cm$^{-2}$, the intervening 
absorber at $z_{\rm abs}=1.3439$ towards PKS 0215$+$01
and the possibly associated absorber at $z_{\rm abs}=0.9107$  towards PKS 0823$-$223.  
The other properties of DLAs and sub-DLAs mentioned above are their large 
$w_{\rm r}$(Fe\,{\sc ii}2600)/$w_{\rm r}$(Mg\,{\sc ii}2796) ratio ($>$0.5) 
and  Mg\,{\sc i}2852 strength  (e.g. Bergeron \& Stasi\' nka \cite{bergeron86s}; 
Rao et al. \cite{rao}). Four strong Mg\,{\sc ii} absorbers do not satisfy these 
criteria and thus cannot be DLAs: (i) two intervening ones at $z_{\rm abs}=0.7753$ 
in SDSS J024156.4$+$004351 and $z_{\rm abs}=1.4247$ in SDSS J094827.0$+$083940, 
(ii) two possibly associated systems at $z_{\rm abs}=1.0851$ in PKS 1008$+$013 
and $z_{\rm abs}=1.6832$ in SDSS J141927.4$+$044513.  

\begin{figure*}[hb]	
   \centering	
      \includegraphics[height=26.cm,angle=-90.]{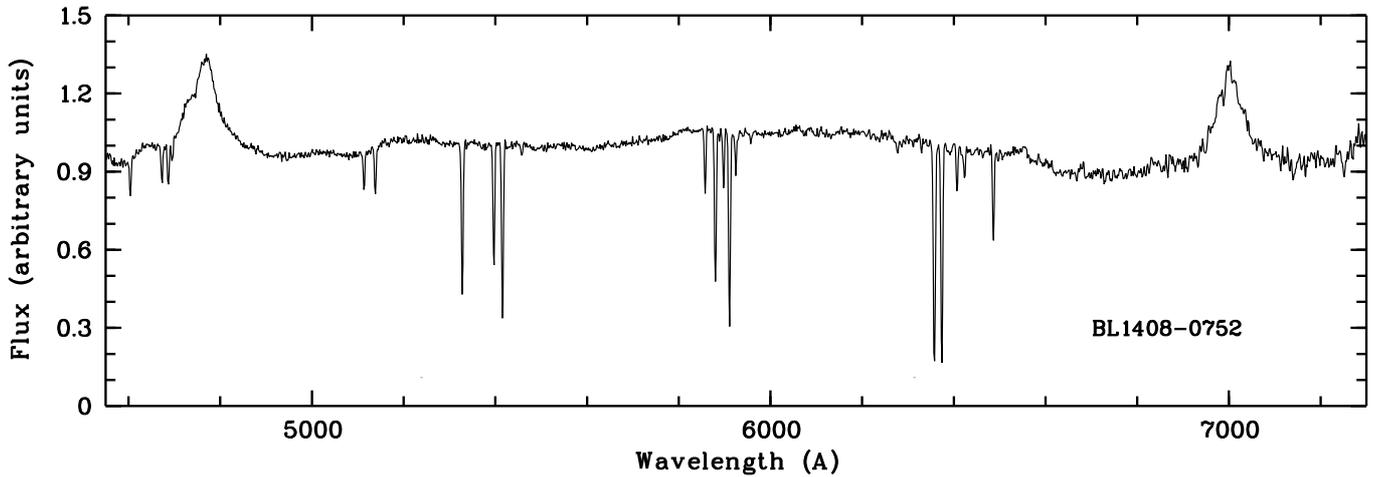}		
    \caption{This Blazar (PKS 1406$-$076) is a variable FSRQ, 
      with superluminal motions, and also a gamma-ray emitter (see Sect. \ref{target}
      and Table \ref{obs}). It has strong emission lines (Mg\,{\sc ii} and C\,{\sc iii}])
      and two Mg\,{\sc ii} absorption systems: 
      a strong one at $z_{\rm abs}= 1.2735$, and a very weak one at $z_{\rm abs}= 1.2913$ 
      thus not included in our statistical Mg\,{\sc ii} sample. 
      The former is a  21~cm absorber and a DLA since, in addition to Mg\,{\sc i}
      and the five stronger Fe\,{\sc ii} lines,  the very weak transitions of 
      Fe\,{\sc ii}2249,2260, Zn\,{\sc ii}2026,2062blend and 
      Cr\,{\sc ii}2056,2062blend,2066 are detected.
}
         \label{BL1408}	
   \end{figure*}

Two strong Mg\,{\sc ii} systems in the BL Lac sample of Stocke \&  Rector 
(\cite{stocke}), also in our sample, were known 21~cm intervening absorbers, at 
$z_{\rm abs}=0.5245$ (DLA mentioned above) and $z_{\rm abs}=0.6850$ towards the lensed 
source B2 0218$+$357. There are three other intervening, strong Mg\,{\sc ii} absorbers 
towards bright radio sources, with DLA signatures (detection of weak transitions of 
Fe\,{\sc ii}, Zn\,{\sc ii} and Cr\,{\sc ii}). We have recently obtained 
GMRT and GBT time to observe these sources. The GMRT data show 21~cm absorption at 
$z_{\rm abs}=1.2753$ in PKS 1406$-$076 and $z_{\rm abs}=1.1158$ in PKS 2029$+$121 
(Gupta et al. in preparation). The FORS1 spectrum of PKS 1406$-$076 is presented in 
Fig. \ref{BL1408}. Analysis of the GBT data for PKS 0454$-$234 is in progress to search 
for 21~cm absorption at $z_{\rm abs}=0.8922$. 

In our Blazar sample, 10 strong Mg\,{\sc ii} systems satisfy the criteria of high 
N(H\,{\sc i}): four are confirmed DLAs and one is a probable DLA, thus yielding a 
DLA fraction of 40\% compatible with that determined for QSOs absorbers. 
Additional UV and/or 21~cm data are needed to ascertain whether the incidence of 
DLA in our strong Mg\,{\sc ii}-Fe\,{\sc ii} Blazar sample is even higher or not 
than for QSOs.  

%

	
\end{appendix}
\end{document}